\documentclass[journal]{IEEEtran}
\usepackage{blindtext}
\usepackage{graphicx}
\usepackage[utf8]{inputenc}
\usepackage{graphics} 
\usepackage{epsfig} 
\usepackage{mathptmx}
\usepackage{amsmath} 
\usepackage{amssymb}  
\usepackage{cite}
\usepackage{multicol}
\usepackage{mathtools, cuted}
\usepackage[cmintegrals]{newtxmath}
\usepackage{epstopdf}
\usepackage{graphics} 
\usepackage{epsfig} 
\usepackage{mathptmx} 
\usepackage{amsmath} 
\usepackage{amssymb}  
\usepackage{cite}
\usepackage{multicol}
\usepackage{mathtools, cuted}
\usepackage[cmintegrals]{newtxmath}
\usepackage{mathtools}

\usepackage{mathptmx}
\usepackage{helvet}
\usepackage{courier}
\usepackage{cancel}
\usepackage{subfigure}
\usepackage{type1cm}
\usepackage{epsfig} 
\usepackage{mathptmx} 
\usepackage{amsmath} 
\usepackage{amssymb}  
\usepackage{cite}
\usepackage{multicol}
\usepackage{mathtools, cuted}
\usepackage[cmintegrals]{newtxmath}
\usepackage{makeidx}         
\newtheorem{theorem}{Theorem}[section]

\usepackage{algorithmic}

\usepackage{multicol}        
\usepackage[bottom]{footmisc}
\usepackage{nomencl}
\usepackage[usenames, dvipsnames]{color}
\newtheorem{prop}{Proposition}

\ifCLASSINFOpdf
\else
\fi

\begin{document}
%
\title{Physics-Based Freely Scalable Continuum  Deformation  
for UAS Traffic Coordination}
%
%
%

\author{Hossein~Rastgoftar and Ella Atkins
\thanks{Hossein Rastgoftar is with the Department
of Aerospace Engineering, University of Michigan, Ann Arbor,
MI, 48109 USA e-mail: hosseinr@umich.edu (see http://a2sys.engin.umich.edu/people/hossein-rastgoftar/).}
\thanks{Ella M. Atkins is with the Department
of Aerospace Engineering, University of Michigan, Ann Arbor,
MI, 48109 USA e-mail: ematkins@umich.edu (see https://aero.engin.umich.edu/people/ella-atkins/).}
}
\maketitle

\begin{abstract}
This paper develops a novel physics-inspired traffic coordination approach and applies it to Unmanned Aircraft System (UAS) traffic management. We extend available physics-inspired approaches previously applied to 1-D traffic flow on highways and urban streets to support models of traffic coordination in higher dimension airspace for cases where no predefined paths exist. The paper considers airspace as a
finite control volume while UAS coordination, treated as continuum deformation, is controlled at the airspace boundaries. By partitioning airspace into planned and unplanned spaces, the paper models nominal coordination in the planned airspace as the solution of a partial differential equation with spatiotemporal parameters.  This paper also improves resilience to vehicle failures with a resilient boundary control algorithm to update the geometry of the planned space when UAS problems threaten safe coordination in existing navigable airspace channels. To support UAS coordination at the microscopic level, we propose clustering vehicles based on vehicle performance limits. UAS clusters, with each UAS treated as a particle of a virtual rigid body, use leader-follower containment to acquire the macroscopic desired trajectory.

\end{abstract}



%
\IEEEpeerreviewmaketitle

\section{Introduction}
Multi-agent coordination has been widely investigated over the past two decades.
Centralized and decentralized coordination approaches typically assume the total number of agents is fixed in a given motion space. This assumption must be relaxed in emerging multi-agent coordination applications such as Unmanned Aerial System (UAS) Traffic Management (UTM) in which the capacity of a finite airspace volume or sector must vary over time, particularly to achieve the high-density operational throughput levels anticipated for urban regions. This paper introduces the notion of \textit{free scalability} for UTM in which the number of vehicles and the specific coordination solution are governed by controlling vehicle inflow and outflow at the boundaries of a motion finite space.  This macroscopic modeling approach is inspired by the enroute airspace sector used for commercial aviation today adapted to an autonomous UTM paradigm of the future.

This paper models air traffic management as \textit{multi-agent coordination}, a topic widely studied in the literature. 
Formation and cooperative control of a multi-agent system can enhance resilience to failure \cite{rieger2013resilient}, improve efficiency, and reduce mission cost \cite{zhao2013energy}. 
Applications include surveillance \cite{botts2016multi}, air traffic  management \cite{zhu2015junction}, formation flight \cite{oh2015survey}, and connected vehicle control \cite{feng2015real}.  
Virtual structures \cite{ren2004decentralized, li2008formation}, consensus \cite{ren2007information, li2018nonlinear, zhang2010consensus}, containment control \cite{li2016containment, zhao2015finite,  notarstefano2011containment}, partial-differential equation (PDE) based approaches \cite{kim2008pde, frihauf2011leader, ghods2012multiagent}, continuum deformation \cite{rastgoftar2016continuum, rastgoftar2017continuum}, and graph rigidity methods \cite{wang2018distance, zhao2018affine} are available multi-agent  coordination approaches. While the virtual structure is a centralized approach, the others are typically implemented in a decentralized fashion. In addition, containment control, PDE-based, and continuum deformation methods apply leader-follower coordination in which leaders move independently and followers acquire the desired coordination through strictly local communication.

The above coordination approaches assume the total number of agents is constant over time within a designated volume or sector. This is a limiting assumption for a traffic system in which vehicles can dynamically leave or enter a traffic control sector. This limitation can be addressed by modeling traffic flow as a spatiotemporal system governed by a partial differential equation \cite{aw2000resurrection, bressan2015conservation, delle2014scalar, pfaff2015optimal}. Ref. \cite{bressan2015conservation} uses mass conversation to model traffic flow near intersections. Ref. \cite{krishnan2018distributed} models swarm coordination as an optimal mass transport problem governed by the continuity PDE.  A coupled PDE-ODE approach is developed in \cite{delle2014scalar} to analyze the underlying influence of large and slow vehicles on large-scale traffic for highways. Multi-lane traffic flow is modeled by a first-order mass conservation PDE in Ref. \cite{shiomi2015multilane}.
In Refs. \cite{como2016convexity, jiang2018macroscopic, pfaff2015optimal} optimal control is integrated with  physics-based models to optimize highway traffic flow. 

Available physically-inspired approaches mostly study 1-D traffic flow in highways and urban areas with mapped roads, pathways, and highways. In contrast, this paper develops an approach to low-altitude UAS coordination (LAUC) in a free airspace with no predefined paths. Autonomous collision avoidance is one of the main LAUC challenges. Each UAS must have sense and avoid (SAA) \cite{kang2017sense, zhu2015vision, ramasamy2015unified, sabatini2014laser, lyu2015vision} capabilities to avoid collision. SAA using laser/lidar \cite{sabatini2014laser} and machine vision \cite{ramasamy2016lidar} technologies is widely studied in the literature. Ref. \cite{kang2017sense} develops a vision-based SAA algorithm to avoid collision through trajectory prediction. SAA using Boolean decision logic (BDL) is proposed in \cite{ramasamy2015unified}. SAA using geofence boundary violation detection and turn-back have been proposed in Refs. \cite{stevens2018geofencing, stevens2018geofence, stevens2016multi} to avoid collision by satisfying keep-in airspace permission constraints.

\subsection{Contribution and Outline}
This paper develops a continuum deformation framework for traffic coordination (CDFTC) in a finite motion space and applies CDFTC to LAUC. 
CDFTC applies principles of continuum mechanics \cite{cueto2018introduction} to achieve freely scalable UAS coordination in a finite airspace volume.  CDFTC integrates the following two main layers: (i) Macroscopic coordination using Eulerian continuum mechanics and (ii) Microscopic coordination of clustered vehicles treated as particles of a virtual rigid body.  
By dividing the motion space into planned and unplanned spaces, the macroscopic layer assigns UAS nominal coordination over the \underline{planned space} in a centralized fashion. To this end, macroscopic coordination is defined as the solution of a parabolic spatiotemporal partial differential equation (PDE).
In the microscopic layer, CDFTC applied to LAUC allows a UAS to move either individually or as part of a group. By treating agents as particles of a continuum (or deformable body), a leader-follower containment approach is developed to coordinate clusters of UAS in a decentralized fashion.

This paper applies the principles of continuum mechanics to manage UAS coordination. To our knowledge, this is the first paper to propose a physics-based macroscopic or \textit{meta-level} coordination law that assures each UAS cluster can safely transit local planned airspace channels via decentralized containment control. By using an Eulerian description of continuum mechanics, time and space allocated to each UAS can be efficiently managed. This paper also supports UAS heterogeneity in both macroscopic and microscopic levels whereas most of the cooperative control literature assumes all vehicles have the same performance. By shaping the spatiotemporal parameters of the macroscopic coordination PDE, we show how UAS with different nominal operation speeds can efficiently and safely occupy a finite airspace volume or sector. At the microscopic level, the paper treats UAS as particles of $0$-D (single UAS), $1$-D, $2$-D, and $3$-D virtual rigid bodies as appropriate to acquire the desired macroscopic coordination in a decentralized fashion. For this purpose, we apply leader-follower containment theory.    

This paper is organized as follows.  A problem statement is presented in Section \ref{Problem Statement}. Functionality of the CDFTC macroscopic and microscopic layers is described in Sections \ref{Macroscopic Coordination} and \ref{Microscopic Coordination}, respectively. Simulation results presented in Section \ref{Simulation Results} are followed by concluding remarks in Section \ref{Conclusion}.


\section{Problem Statement}
\label{Problem Statement}
\begin{figure}[ht]
\center
\includegraphics[width=3.3 in]{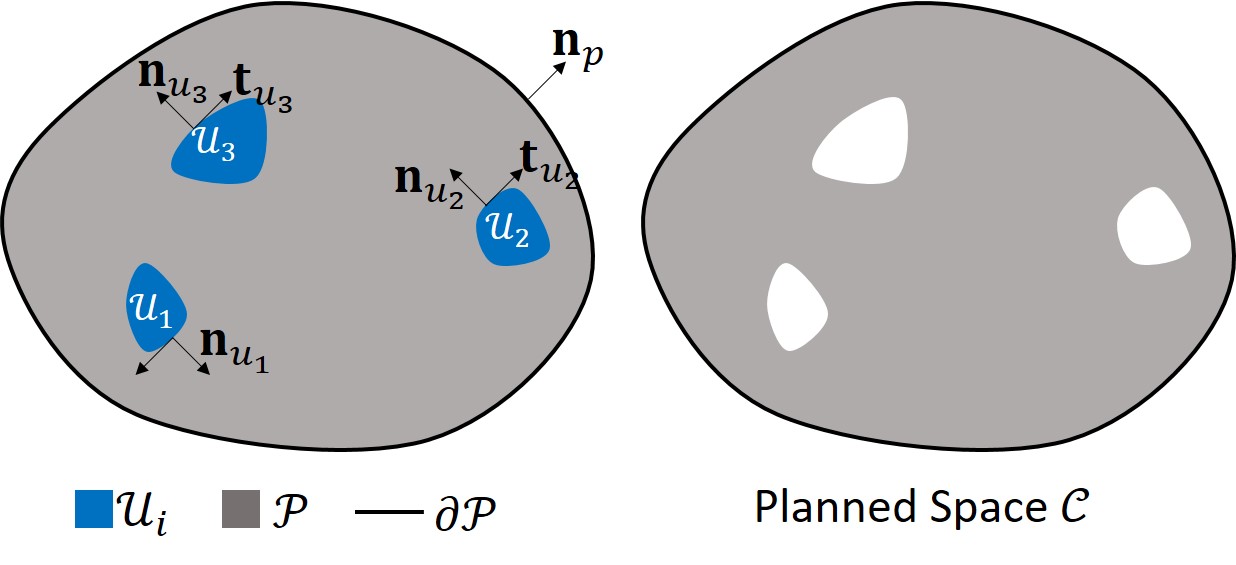}
\caption{Planned-unplanned partitioning of the airspace.}
\label{CompleteMotionSpace1}
\end{figure}

The paper applies principles of continuum mechanics to safely coordinate a large-scale MAS in a finite airspace.  A finite airspace sector requiring LAUC is defined by open set $\mathcal{P}$
with boundary $\partial \mathcal{P}$. Assume that $\mathcal{P}$ contains a finite number of obstacles defined by closed set unplanned set $\mathcal{U}$.
Then,
\[
\mathcal{C}=\left(\partial \mathcal{P}\bigcup\mathcal{P}\right)\setminus \mathcal{U}
\]
defines the planned subset of the airspace.
Note that $\mathcal{U}=\{\mathcal{U}_1,\cdots,\mathcal{U}_{n_u}\}$ defines obstacle zones $n_u$ excluded from $\mathcal{P}$. A schematic of planned-unplanned airspace partitioning is shown in Fig. \ref{CompleteMotionSpace1}. {\color{black}As shown in Fig. \ref{CompleteMotionSpace1},  $\mathbf{n}_{u_i}(\mathbf{r},t)$ ($i=1,\cdots,n_u$) denotes the outward unit normal vector on the unplanned airspace boundary while $\mathbf{n}_{p}(\mathbf{r},t)$ is the outward unit normal vector on airspace boundary $\partial\mathcal{P}$. Furthermore,  $\mathbf{t}_{u_i}(\mathbf{r},t)$ ($i=1,\cdots,n_u$) is the normal tangent vector at unplanned boundary $\partial \mathcal{U}_i$ ($i=1,\cdots,n_u$).}

For macroscopic coordination, we use an Eulerian description of continuum mechanics to define nominal coordination over the planned airspace and manage time interval and space allocation to UAS across the planned airspace in a computationally-effective manner. In practice, this maximizes airspace usability by {\color{black} scaling services provided to UAS while they exist in the planned airspace.} This paper models UAS macroscopic coordination as the solution of a PDE with spatiotemporal parameters. To enhance resilience of the macroscopic coordination against failure in the planned airspace, we offer a novel boundary control approach to update geometry of the planned airspace in a safe manner.

For microscopic coordination, UAS clusters are formed based on identifying compatible nominal operational speeds of all aerial vehicles operating in the region. UAS clusters are then treated as a particles of virtual rigid bodies with 3-D rotations and translations in the planned airspace. A leader-follower containment approach is used by each cluster to track the nominal coordination assigned at the macroscopic level in a decentralized fashion.

\section{Macroscopic Coordination}
\label{Macroscopic Coordination}
For macroscopic coordination, an Eulerian description of continuum mechanics is used to define a spatio-temporal UAS cluster velocity field over $\mathcal{C}$, represented as  $\mathbf{V}=\mathbf{V}\left(\mathbf{r},t\right)$ where $\mathbf{r}$ and $t$ are three-dimensional position and time, respectively. Defining a Cartesian coordinate system with unit bases $\hat{\mathbf{e}}_x$, $\hat{\mathbf{e}}_y$, and $\hat{\mathbf{e}}_z$, position and velocity are expressed as follows:
\[
\begin{split}
    \mathbf{r}=&x\hat{\mathbf{e}}_x+y\hat{\mathbf{e}}_y+z\hat{\mathbf{e}}_z\\
    \mathbf{V}=&v_x\hat{\mathbf{e}}_x+v_y\hat{\mathbf{e}}_y+v_z\hat{\mathbf{e}}_z\\
\end{split}
,
\]
where $x$, $y$, and $z$ are position components and $v_x$, $v_y$, and $v_z$ are velocity components.

\textit{\underline{Governing Equation:}} Defining an (actual) potential function $\Phi_a=\Phi_a(\mathbf{r},t)$ where $\mathbf{r}\in \mathcal{P}$ denotes position in the planned space and $t$ denotes time. Nominal velocity of the field is given by
\begin{equation}
\label{referencedesiredvelocity}
     \mathbf{r}\in \mathcal{C},~t\geq 0,\qquad \mathbf{V}\left(\mathbf{r},t\right)=K\left(\mathbf{r},t\right)\bigtriangledown \Phi_a\left(\mathbf{r},t\right),
\end{equation}
where $K(\mathbf{r},t):(\mathcal{C},\mathbb{R}_+)\rightarrow \mathbb{R}_+$ is the cost 
and $\Phi_a(\mathbf{r},t)$ is governed by the following partial differential equation (PDE):
\begin{equation}
\label{goveqdyn}
   \mathbf{r}\in \mathcal{C},~t\geq 0,\qquad \dfrac{{d}\Phi_a}{dt}=\bigtriangledown\cdot\left(\bigtriangledown\Phi_a\right)=\bigtriangledown\cdot\left(K\left(\mathbf{r},t\right)\mathbf{V}\left(\mathbf{r},t\right) \right),
\end{equation}
where $\bigtriangledown=\left[\frac{\partial}{\partial x}~\frac{\partial}{\partial y}~\frac{\partial}{\partial z}\right]^T$ is the gradient vector defined with respect to the Cartesian coordinate system. 




\subsection{Reference Macroscopic Coordination}
Let $\Phi=\Phi(\mathbf{r},t)$ represent a reference potential field. A reference macroscopic coordination is the solution of 
\begin{equation}
    \label{goveq}
    \bigtriangledown\left(\bigtriangledown \Phi\right)=0
\end{equation}
This paper applies the divergence theorem to relate the volume integral $\bigtriangledown\cdot\left(\bigtriangledown \Phi\right)$ over $\mathcal{C}\setminus \left(\partial \mathcal{U}\bigcup\partial \mathcal{P}\right)$ and the surface integral of $\bigtriangledown \Phi$ over the borders $\partial \mathcal{U}$ and $\partial \mathcal{P}$ by
\begin{equation}
\label{DIVERGENCE1}
\begin{split}
    &\int_{\mathcal{C}\setminus \left(\partial \mathcal{U}\bigcup\partial \mathcal{P}\right)}\left(\bigtriangledown\cdot\left(\bigtriangledown \Phi_a\right)\right)\mathrm{d}V=\int_{\partial \mathcal{P}}\left(\bigtriangledown \Phi_a\right)\cdot \hat{\mathbf{n}}_{p}\mathrm{d}A\\
    -& \sum_{j=1}^{n_u}\left(\int_{\partial \mathcal{U}_j}\cancelto{0}{\left(\bigtriangledown \Phi_a\right)\cdot \hat{\mathbf{n}}_{u_j}}\mathrm{d}A\right)=\int_{\partial \mathcal{P}}\left(\bigtriangledown \Phi_a\right)\cdot \hat{\mathbf{n}}_{p}\mathrm{d}A.
\end{split}
\end{equation}
In \eqref{DIVERGENCE1}, $\partial \mathcal{U}_j$ defines a boundary surface enclosing the unplanned airspace $\mathcal{U}_j$, $\partial \mathcal{U}=\{\partial\mathcal{U}_{1},\cdots,\partial\mathcal{U}_{n_u}\}$, $\mathbf{n}_{u_j}$ is the outward unit vector normal to the boundary surface $\partial\mathcal{U}_j$, and $\mathbf{n}_{p}$ is the outward unit vector normal to the boundary of the airspace. 
Note that $\left(\bigtriangledown \Phi_a\right)\cdot \hat{\mathbf{n}}_{o_j}$ vanishes because normal velocity component is zero at any point on the obstacle boundaries. Define boundary flux
\begin{equation}
\label{boundEqualConts}
\mathbf{r}\in \partial \mathcal{P},\qquad q\left(\mathbf{r},t\right)=\bigtriangledown\Phi\cdot \hat{\mathbf{n}}_p.
\end{equation}
A key assumption of this paper is that 
\begin{equation}
\label{BoundaryCoordination}
    \forall t,\qquad \int_{\partial \mathcal{P}}q\left(\mathbf{r},t\right)\mathrm{d}A=\int_{\partial \mathcal{P}}\left(\bigtriangledown \Phi\right)\cdot \hat{\mathbf{n}}_{p}\mathrm{d}A=0.
\end{equation}
To assign a nominal coordination, we assume that the left-hand side of Eq. \eqref{BoundaryCoordination} vanishes at any time $t$. This leads the  nominal macroscopic coordination to be assigned as the solution of the Laplace Equation:
\begin{equation}
\label{LaplaceEquationnnnnn}
    \forall t,~\mathbf{r}\in \mathcal{C},\qquad \bigtriangledown\cdot\left(\bigtriangledown \Phi\right)=\bigtriangledown\cdot\left(\bigtriangledown\cdot( K(\mathbf{r},t)\mathbf{V}(\mathbf{r},t)\right))=0.
\end{equation}

Notice that $\Phi$ defines the steady state solution $\Phi=\Phi(\mathbf{r})$ if: (i)  $q=q(\mathbf{r})$ is spatially varying and time-invariant at any point on the outer boundary $\partial \mathcal{P}$, and (ii) the geometry of unplanned space $\mathcal{U}$ does not change. Otherwise, nominal coordination can vary with time at a point $\mathbf{r}\in \mathcal{C}$, but $\Phi$ is still a solution of Laplace equation \eqref{goveq}. In other words, both spatially-varying and spatiotemporally-varying solutions of nominal (reference) coordination $\Phi$ satisfy the Laplace Equation.

Next, Section \ref{Macroscopic Coordination} develops numerical and analytic solutions to the macroscopic coordination governing PDE. Microscopic coordination of each UAV/agent cluster is then discussed in Section \ref{Microscopic Coordination}.
Collectively these formulations offer numerical and analytic approaches to solve  the governing coordination  equation given in \eqref{LaplaceEquationnnnnn}.


\subsubsection{Numerical Solution}
\label{Nominal Velocity Assignment}
This paper applies the finite difference (FD) method  to assign $\Phi$ over $\mathcal{C}$. 
Let a connected network with $m$ nodes, defined by $\mathcal{V}=\{1,2,\cdots,m\}$, be distributed over airspace $\mathcal{P}\bigcup \partial \mathcal{P}$. It is assumed that the geometry of airspace outer border $\partial \mathcal{P}$ is fixed. Therefore, the node set $\mathcal{V}$ is time-invariant, e.g.  $\left|\mathcal{V}\right|=m$ is unchanged. Let $\mathcal{V}_{u}=\{i_1,\cdots,i_{m_u}\}$ define those grid nodes distributed over unplanned airspace $\mathcal{U}$ and  $\mathcal{V}_{c}=\mathcal{V}\setminus \mathcal{V}_u$ define the remaining grid nodes distributed over planned airspace $\mathcal{C}$. $\mathcal{V}_c$ can be expressed as $\mathcal{V}_c=\mathcal{V}_{cb}\bigcup\mathcal{V}_{ci}$, where
\begin{enumerate}
    \item{$\mathcal{V}_{cb}=\left\{j_1,\cdots,j_{m_{cb}}\right\}$ defines the outer boundary control nodes distributed over $\partial \mathcal{P}$,} 
    \item{$\mathcal{V}_{ci}=\left\{l_1,\cdots,l_{m_{ci}}\right\}$ defines the interior control nodes distributed over $\mathcal{C}\setminus\left( \partial \mathcal{U}\bigcup \partial \mathcal{P}\right)$.} 
\end{enumerate}

Because unplanned airspace can be potentially dynamic, $m_u(t)=\left|\mathcal{V}_{u}\right|$ can change with time. Therefore,  $m_{ci}(t)=\left|\mathcal{V}_{ci}\right|$ can be time-varying as well. However, $m_{cb}+m_{ci}(t)+m_u(t)=m$ remains unchanged at any time $t$.

Nodal $\Phi$ values, defined by ${\mathbf{\Phi}}=[\Phi_i]\in \mathbb{R}^{m\times1}$ are found by solving 
\begin{equation}
\label{PHIII}
-\mathbf{L}\left(t\right){\mathbf{\Phi}}\left(t\right)+{\mathbf{q}}\left(t\right)=\mathbf{0},
\end{equation}
where $\Phi_i\left(t\right)=\Phi\left(x_i,y_i,z_i,t\right)$ ($i\in \mathcal{V}$), ${\mathbf{q}}\left(t\right)=[q_i]\in \mathbb{R}^{m\times 1}$ is the flux vector, and $\mathbf{L}\left(t\right)\in \mathbb{R}^{m\times m}$ is the positive semi-definite Laplacian matrix.

Define $\hat{{\mathbf{\Phi}}}=\mathbf{S}{\mathbf{q}}=
\begin{bmatrix}
\hat{\mathbf{\Phi}}_{cb}^T&
\hat{\mathbf{\Phi}}_{ci}^T&
\hat{\mathbf{\Phi}}_{u}^T
\end{bmatrix}^T$, 

\noindent $\hat{\mathbf{q}}=\mathbf{S}\mathbf{q}=
\begin{bmatrix}
\hat{\mathbf{q}}_{cb}^T&
\hat{\mathbf{q}}_{ci}^T&
\hat{\mathbf{q}}_{u}^T
\end{bmatrix}^T$, and 
\[
\hat{\mathbf{L}}=\mathbf{S}\mathbf{L}\mathbf{S}^T=
\begin{bmatrix}
\hat{\mathbf{L}}_{cb,cb}&\hat{\mathbf{L}}_{cb,ci}&\hat{\mathbf{L}}_{cb,u}\\
\hat{\mathbf{L}}_{cu,cb}&\hat{\mathbf{L}}_{cu,ci}&\hat{\mathbf{L}}_{cu,u}\\
\hat{\mathbf{L}}_{u,cb}&\hat{\mathbf{L}}_{u,ci}&\hat{\mathbf{L}}_{u,u}\\
\end{bmatrix}
,
\]
where superscript ``$T$'' denotes matrix transpose and orthogonal matrix $\mathbf{S}=[\mathbf{S}_{hq}]\in \mathbb{R}^{m\times m}$, called the \emph{interface matrix}, is given by
\begin{equation}
    \mathbf{S}_{hq}=
    \begin{cases}
    1&h=1,\cdots,m_{cb}\wedge\left(j_h\in { \mathcal{V}}_{cb}\right)\wedge j_h=q\\
    1&h=m_{cb}+1,\cdots,m_{cb}+m_{ci}\wedge\left(l_h\in { \mathcal{V}}_{ci}\right)\wedge l_h=q\\
    1&h=m_{cb}+m_{ci}+1,\cdots,m\wedge\left(i_h\in { \mathcal{V}}_{u}\right)\wedge i_h=q\\
    0&\mathrm{else}.
    \end{cases}
\end{equation}
Note that $\hat{\mathbf{q}}_{ci}=\mathbf{0}\in \mathbb{R}^{m_{ci}\times 1}$ and $\hat{\mathbf{q}}_{u}=\mathbf{0}\in \mathbb{R}^{m_{u}\times 1}$. 
Let $\mathbf{S}(t)=\begin{bmatrix}\mathbf{S}_{cb}\\ \mathbf{S}_{cu}(t)\\\mathbf{S}_{ci}(t)\end{bmatrix}$, where $\mathbf{S}_{cb}\in \mathbb{R}^{m_{cb}\times m}$, $\mathbf{S}_{cu}(t)\in \mathbb{R}^{m_{cu}\times m}$, and $\mathbf{S}_{ci}(t)\in \mathbb{R}^{m_{ci}\times m}$. 
By substituting $\mathbf{L}=\mathbf{S}^T\hat{\mathbf{L}}{\mathbf{S}}$, $\mathbf{q}=\mathbf{S}^T\hat{\mathbf{q}}$, $\mathbf{{\mathbf \Phi}}=\mathbf{S}^T\hat{\mathbf{{\mathbf{\Phi}}}}$, Eq. \eqref{PHIII} is converted to $\hat{\mathbf{L}}\hat{{\Phi}}=\hat{{\mathbf{q}}}$, or
\begin{equation}
\label{trasformedlaplace}
-\hat{\mathbf{L}}(t)\hat{{\Phi}}(t)=\hat{{\mathbf{q}}}(t).
\end{equation}
We assume that $\hat{\mathbf{\Phi}}_{cb}$ is given, e.g. $\Phi$ values of the outer boundary nodes are not impacted by the $\Phi$ values of the in-neighbor nodes. Therefore,  $\hat{\mathbf{L}}_{cb,cu}=\mathbf{0}$, $\hat{\mathbf{L}}_{cb,ci}=\mathbf{0}$, and $\hat{\mathbf{L}}_{cb,u}=\mathbf{0}$. Furthermore, $\hat{\mathbf{\Phi}}_u=\mathbf{0}$ at any time $t$. 
 Therefore, the unplanned nodes defined by $\mathcal{V}_{u}$ are independent:  $\hat{\mathbf{L}}_{u,cb}=\mathbf{0}$, $\hat{\mathbf{L}}_{u,ci}=\mathbf{0}$, and $\hat{\mathbf{L}}_{u,cu}=\mathbf{0}$. Consequently, 
\begin{equation}
\left[
\begin{array}{c|c}
-\hat{\mathbf{L}}_{cb,cb}&\mathbf{0}\\
\hline
\mathbf{B}_c&\mathbf{A_c}
\end{array}
\right]
=-\left[
\begin{array}{c|ccc}
\hat{\mathbf{L}}_{cb,cb}&\mathbf{0}&\mathbf{0}\\
\hline
\hat{\mathbf{L}}_{ci,cb}&\hat{\mathbf{L}}_{ci,ci}&\hat{\mathbf{L}}_{ci,u}\\
\mathbf{0}&\mathbf{0}&\mathbf{I}_{m_u}\\
\end{array}
\right]
\end{equation}
where $\hat{\mathbf{L}}_{cb,cb}$ is \underline{diagonal and positive definite},  $\mathbf{I}_{m_u}\in \mathbb{R}^{m_u\times m_u}$ is the identity matrix, and $\mathbf{A}_c$ has the following properties: \begin{enumerate}
    \item{Diagonal entries of matrix $\mathbf{A}_c$ are all negative.}
    \item{Off-diagonal entries of matrix $\mathbf{A}_c$ are all non-negative.} 
    \item{The sum of the entries are non positive for every row of matrix $\mathbf{A}_c$.}
\end{enumerate}

\begin{theorem}\label{theorem1}
{\color{black}If the network distributed over the airspace is connected, matrix $\mathbf{A}_c$ is Hurwitz.}
\end{theorem}

Steady-state coordination Eq. \eqref{PHIII} can be rewritten as
\begin{equation}
\label{DESIREEEE}
    \mathbf{A}_{c}\mathbf{\Phi}_I
    +\mathbf{B}_{c}\hat{\mathbf{\Phi}}_{cb}=\mathbf{0},
\end{equation}

where $\mathbf{\Phi}_I=\begin{bmatrix}
\hat{\mathbf{\Phi}}_{ci};
\hat{\mathbf{\Phi}}_{u}
\end{bmatrix}\in \mathbb{R}^{\left(m-m_{cb}\right)\times 1}$.


\begin{figure}[ht]
\centering
\includegraphics[width=0.4\textwidth]{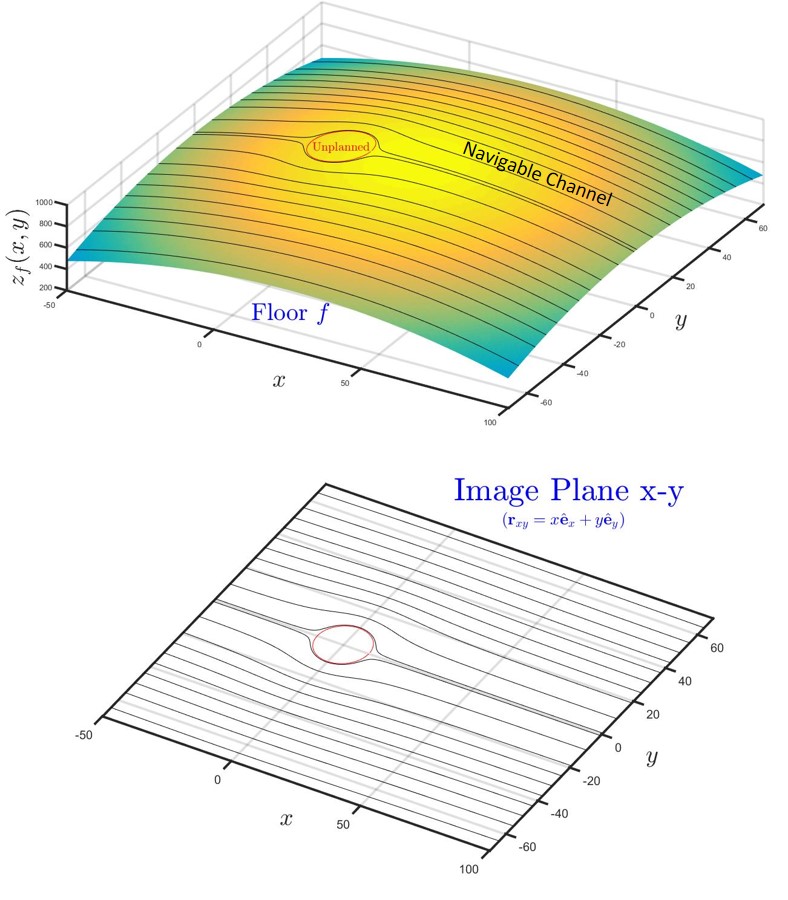}
\caption{\label{analytic}Schematic of floor $f$ defined as the analytic solution of governing Eq. \eqref{LaplaceEquationnnnnn}.}
\end{figure}

\subsubsection{Analytic Solution}
\label{Analytic Solution}
For the analytic solution, the airspace is decomposed into $n_s$ motion surfaces (or floors) defined by 
$\Gamma(x,y,z,t)=\Gamma_f$ where $\Gamma_f$ is a constant for each $f=1,\cdots,n_s$. A position on floor $f$ can be expressed as  $\mathbf{r}_f=\mathbf{r}\left(\Phi_f,\Psi_f,\Gamma_f,t\right)$, where $\Phi_f=\Phi_f\left(x,y,t\right)$ is the potential function of floor $f$ satisfying  the governing PDE in \eqref{goveq},  and $\Psi_f=\Psi_f\left(x,y,t\right)$ is a stream function.

\textbf{Remark 1:} The level curves $\Phi_f\left(x,y,t\right)=\rm{constant}$ and $\Psi_f\left(x,y,t\right)=\rm{constant}$ perpendicularly cross at the intersection point in the image plane $x-y$. Using the Cauchy Riemann theorem \cite{kreyszig2010advanced}, $\Psi_f$ and $\Phi_f$ are related as follows: 
\begin{subequations}
\begin{equation}
    \frac{\partial\Phi_f}{\partial x}=\frac{\partial\Psi_f}{\partial y}=K_f(x,y,t)u_f(x,y,t)
\end{equation}
\begin{equation}
    \frac{\partial\Phi_f}{\partial y}=-\frac{\partial\Psi_f}{\partial x}=K_f(x,y,t)v_f(x,y,t),
\end{equation}
\end{subequations}
where $u_f$ and $v_f$ are the nominal velocity components on floor $f$ projectes on the $x-y$ plane. Note that $\Phi_f$ and $\Psi_f$ both satisfy Laplace equations $\bigtriangledown^2\Phi_f=0$ and  $\bigtriangledown^2\Psi_f=0$.


\begin{table}
    \centering
    \caption{2-D ideal fluid flow Potential functions}
    \begin{tabular}{|c|c|}
          \hline
    Pattern& Potential Function  \\
    \hline
     Uniform&${\Phi}_{\rm{U}}\left(\mathbf{r}_{xy},\mathbf{U}_{\infty,f}(t)\right)=\mathbf{U}_{\infty,f}\cdot \mathbf{r}_{xy}$\\
     \hline
   Source&${\Phi}_{\rm{So}}\left(\mathbf{r}_{xy},\mathbf{r}_{0,i,f}\right)=\Delta_i\ln{\|\mathbf{r}_{xy}-\mathbf{r}_{0,i,f}\|}$\\
   \hline
   Sink&${\Phi}_{\rm{Si}}\left(\mathbf{r}_{xy},\mathbf{r}_{0,i,f}\right)=-\Delta_i\ln{\|\mathbf{r}_{xy}-\mathbf{r}_{0,i,f}\|}$\\
   \hline
   Doublet&${\Phi}_{\rm{D}}\left(\mathbf{r}_{xy},\mathbf{r}_{0,i,f}\right)= \frac{\Delta_i\left(\mathbf{r}_{xy}-\mathbf{r}_{0,i,f}\right)\cdot\mathbf{n}_{f}}{\|\mathbf{r}_{xy}-\mathbf{r}_{0,i,f}\|}$\\
   \hline    
    \end{tabular}
      \label{Table:1}
\end{table}
\begin{table}
    \centering
    \caption{2-D ideal fluid flow Potential functions}
    \begin{tabular}{|c|c|}
          \hline
    Pattern & Stream Function  \\
    \hline
     Uniform&${\Psi}_{\rm{U}}\left(\mathbf{r}_{xy},\mathbf{U}_{\infty,f}(t)\right)=u_{\infty,f}\left(y\cos\theta_{0,f}-x\sin\theta_{0,f}\right)$\\
     \hline
   Source&${\Psi}_{\rm{So}}\left(\mathbf{r}_{xy},\mathbf{r}_{0,i,f}\right)=\Delta_i\tan^{-1}\left(\dfrac{y-y_{0,i,f}}{x-x_{0,i,f}}\right) $\\
   \hline
   Sink&${\Psi}_{\rm{Si}}\left(\mathbf{r}_{xy},\mathbf{r}_{0,i,f}\right)=-\Delta_i\tan^{-1}\left(\dfrac{y-y_{0,i,f}}{x-x_{0,i,f}}\right) $\\
   \hline
   Doublet&${\Psi}_{\rm{D}}\left(\mathbf{r}_{xy},\mathbf{r}_{0,i,f}\right)= \frac{-\Delta_i\left[\sin\theta_{0,f}\left(x-x_{0,i,f}\right)+\cos\theta_{0,f}\left(y-y_{0,i,f}\right)\right] }{ \left(x-x_{0,i,f}\right)^2+\left(y-y_{0,i,f}\right)^2}$\\
   \hline    
    \end{tabular}
      \label{Table:1a}
\end{table}

Let position vector $\mathbf{r}_f$ be expressed as 
\[
\mathbf{r}_f=\mathbf{r}_{xy}+z_f(x,y)\hat{\mathbf{e}}_z
\]
where 
$\mathbf{r}_{xy}=x\hat{\mathbf{e}}_x+y\hat{\mathbf{e}}_y$ is the projection of $\mathbf{r}$ on the $x-y$ plane (image plane), and $\hat{\mathbf{e}}_x$, $\hat{\mathbf{e}}_y$, and $\hat{\mathbf{e}}_z$ are the bases of the Cartesian coordinate system. 
Macroscopic coordination is treated as ideal fluid flow or irrotational flow \cite{munson2013fluid} defined in the $x-y$ image plane per the Fig. \ref{analytic} schematic.  ``Uniform Flow'', ``Source'', ``Sink'', and ``Doublet'' are ideal flow patterns considered in this paper. Potential and stream functions $\Phi$ and $\Psi$ of these ideal flow patterns are given in Table \ref{Table:1}.  
Define unplanned airspace characteristic point $\mathbf{r}_{0,i,f}=[x_{0,i,f}~y_{0,i,f}]\in \mathcal{U}_i$ ($i=1,\cdots,n_u$).  Unplanned airspace $\mathcal{U}_i$ can be excluded from $\mathcal{P}$ by combining either a uniform flow and doublet or a uniform flow, a sink and a source. Mathematically speaking, we define
\begin{subequations}
\begin{equation}
\begin{split}
   \Phi_f=&\Phi_{\rm{U}}\left(\mathbf{r}_{xy},\mathbf{U}_{\infty,f}(t)\right)+
   \Xi_f\sum_{i=1}^{n_u}[\gamma_{i,f}\Phi_{\rm{D}}\left(\mathbf{r}_{xy},\mathbf{r}_{0,i,f}\left(t\right)\right)\\
   +&\left(\Phi_{\rm{Si}}\left(\mathbf{r}_{xy},\mathbf{r}_{0,i,f}\left(t\right)\right)+\Phi_{\rm{So}}\left(\mathbf{r},\mathbf{r}_{0,i,f}\left(t\right)\right)\right)]
\end{split}
\end{equation}
\begin{equation}
\begin{split}
    \Psi_f=&\Psi_{\rm{U}}\left(\mathbf{r}_{x,y},\mathbf{U}_{\infty,f}(t)\right)+\Xi_f\sum_{i=1}^{n_u} [\gamma_{i,f} \Psi_{\rm{D}}\left(\mathbf{r}_{xy},\mathbf{r}_{0,i,f}\right)\\+&\left(1-\gamma_{i,f}\right)\left(\Psi_{\rm{Si}}\left(\mathbf{r}_{xy},\mathbf{r}_{0,i,f}\right)+\Psi_{\rm{So}}\left(\mathbf{r}_{xy},\mathbf{r}_{0,i,f}\right)\right)],
\end{split}
\end{equation}
\end{subequations}
where $\mathbf{U}_{\infty,f}=u_{\infty,f}\mathbf{n}_{f}$ is the traffic flow direction projected on the horizontal $x-y$ plane and $\mathbf{n}_{f}(t)= 
 \cos\theta_{0,f}(t)\hat{\mathbf{e}}_x+\sin\theta_{0,f}(t)\hat{\mathbf{e}}_y
 $ is a unit vector. $\gamma_{i,f}$ and $\Xi_f$ are binary parameters with values $0$ or $1$. If floor $f$ contains \underline{no} unplanned space, then $\Xi_f=0$. Otherwise, $\Xi_f=1$. If $\gamma_{i,f}=1$, unplanned airspace $\mathcal{U}_i$ is excluded by combining uniform flow and doublet flow; otherwise, $\mathcal{U}_i$ is excluded by combining uniform flow, a sink, and a source.

 The boundary of the unplanned airspace is defined by
\begin{equation}
    \partial \mathcal{U}=\left\{\mathbf{r}\in \mathcal{P}\big|\Psi_{f}\left(\mathbf{r}_{xy},t\right)=0,~f=1,\cdots,n_s\right\}.
\end{equation}
In the example shown in Fig. \ref{analytic},  unplanned airspace $\mathcal{U}$ is separated from the planned space by combining uniform and doublet flows. Assuming $n_s=1$, $n_u=1$, $\theta_{0,f}=0$, and $\mathbf{r}_{1,f}=\mathbf{0}$, 
\begin{equation}
\label{PhIFFF}
    \phi_f=u_{\infty,f}\left(1+\dfrac{\Delta_f}{x^2+y^2}\right)x
\end{equation}
\begin{equation}
\label{PSIFFF}
    \psi_f=u_{\infty,f}\left(1-\dfrac{\Delta_f}{x^2+y^2}\right)y
\end{equation}
define the potential and stream fields over  projection plane $x-y$, respectively. $\Psi_f=0$ defines the boundary of the unplanned airspace excluding an obstacle by a circle with radius $\sqrt{\frac{\Delta_f}{ u_{\infty,f}}}$. 

\textbf{Remark 2:} $u_{\infty,f}$, $\Delta_{i,f}$, and $\mathbf{r}_{0,i,f}$ are design parameters to appropriately specify the geometry of unplanned airspace $\mathcal{U}_i\in \mathbf\mathcal{U}$. By increasing $u_{\infty,f}$ the size of the unplanned air space is decreased on floor $f$. On the other hand, the unplanned airspace $\mathcal{U}_i$ is enlarged when $\Delta_i$ is increased. Unplanned airspace $\mathcal{U}_i$ can be displaced in airspace $\mathcal{P}$ by updating $\mathbf{r}_{0,i,f}$. 

\subsection{Resilient Macroscopic Coordination}
\label{ResilientMacroscopicCoordination2}
To realize a robust and efficient macroscopic coordination strategy, it is desirable to minimize the nominal velocity variation imposed by  unplanned airspace displacement and deformation when geometry of the unplanned airspace changes with time. {\color{black}To this end, an analytic boundary control solution will be designed in Section \ref{Analytic Boundary Control} to handle unpredicted (pop-up) unplanned airspace sectors cut out from the planned airspace. Furthermore, a numerical boundary control will be developed in Section \ref{Numerical Boundary Control} to enhance  macroscopic coordination in cases where an unplanned airspace volume can either be  stationary or move predictably over time.}

{\color{black}
\subsubsection{Analytic Boundary Control}\label{Analytic Boundary Control}
We define the resilient macroscopic coordination over the floor $f$ as the solution of the following governing PDE:
\begin{equation}
\label{goveqdynnnn}
   f=1,\cdots,n_s,~\mathbf{r}\in \mathcal{C},~t\geq 0,\qquad \dfrac{{d}\Psi_{f,a}}{dt}=\bigtriangledown\cdot\left(\bigtriangledown\Psi_{f,a}\right)
\end{equation}
subject to the boundary condition
\begin{subequations}
\begin{equation}
\label{bougoveqdyn2}
   f=1,\cdots,n_s,~\mathbf{r}\in\partial \mathcal{P},~t\geq 0,\qquad \dfrac{\partial\Psi_{f,a}}{\partial t}=U_p(\mathbf{r},t)
\end{equation}
\begin{equation}
\label{unplannedgoveqdyn1}
   f=1,\cdots,n_s,~\mathbf{r}\in\partial \mathcal{U},~t\geq 0,\qquad \dfrac{\partial\Psi_{f,a}}{\partial t}=U_u(\mathbf{r},t).
\end{equation}
\end{subequations}
where $\psi_{f,a}$ is the actual stream function defined over planned space $\mathcal{C}$. Note that $U_p(\mathbf{r},t)$ and $U_u(\mathbf{r},t)$ 
are the boundary controls actuated at boundaries $ \mathcal{\partial P}$ and $\partial \mathcal{U}$, respectively.

\begin{prop}\label{prop1}
 Let $\mathbf{n}_{u_j}$ and $\mathbf{t}_{u_j}$ denote normal and tangent unit vectors at unplanned boundary $\partial \mathcal{U}_j$ per Fig. \ref{CompleteMotionSpace1}.
 $\psi_f(\mathbf{r},t)$ is constant at any point $\mathbf{r}$ on the boundary of the planned airspace.
 \end{prop}

 Considering Proposition \ref{prop1}, reference macroscopic coordination can be obtained as the solution 
 \begin{equation}
\label{goveqdynn}
   f=1,\cdots,n_s,~\mathbf{r}\in \mathcal{C},~t\geq 0,\qquad \bigtriangledown\cdot\left(\bigtriangledown\Psi_{f}\right)=0
\end{equation}
subject to boundary condition
\begin{subequations}
\begin{equation}
\label{bougoveqdyn1}
   f=1,\cdots,n_s,~\mathbf{r}\in\partial \mathcal{P},~t\geq 0,\qquad \Psi_{f}=\rm{given},
\end{equation}
\begin{equation}
\label{unplannedgoveqdyn2}
   f=1,\cdots,n_s,~\mathbf{r}\in\partial \mathcal{U},~t\geq 0,\qquad \Psi_{f}=0.
\end{equation}
\end{subequations}

\begin{theorem}\label{theroem2}
Let $E_f(\mathbf{r},t)=\psi_{f,a}(\mathbf{r},t)-\psi_{f}(\mathbf{r},t)$ be the distributed error function defining the difference between actual and reference stream values over the planned space ($\mathbf{r}\in \mathcal{C},~t\geq 0$). $E_f(\mathbf{r},t)$ governed by 
\begin{equation}
\label{ergoveq}
f=1,\cdots,n_s,~\mathbf{r}\in \mathcal{C},~t\geq 0,\qquad \dfrac{{d}E_{f}}{dt}=\bigtriangledown\cdot\left(\bigtriangledown E_{f}\right).
\end{equation}
$ {E_{f,a}}(\mathbf{r},t)$ asymptotically tends to $0$, if the boundary controls $U_p$ and $U_u$ are chosen as follows:
\begin{subequations}
\begin{equation}
\label{up}
    f=1,\cdots,n_s,~\mathbf{r}\in\partial \mathcal{P},~t\geq 0,\qquad U_p(\mathbf{r},t)=\dfrac{\Psi_{f,a}}{dt}-k_p\left(\bigtriangledown E_f\cdot\mathbf{n}_p\right)
\end{equation}
\begin{equation}
\label{uu}
\begin{split}
   &f=1,\cdots,n_s,~j=1,\cdots,n_u,~\mathbf{r}\in\partial \mathcal{U}_j,~t\geq 0,\qquad\qquad\qquad\\
   &U_{u_j}(\mathbf{r},t)=-k_{u_j}\left(\bigtriangledown E_f\cdot \mathbf{n}_{u_j}\right)
\end{split}
\end{equation}
\end{subequations}
where $k_p>0$ and $k_{u_j}>0$ are constant control gains.
\end{theorem}
}

\subsubsection{Numerical Boundary Control}
\label{Numerical Boundary Control}
Discretizing governing equation \eqref{goveqdyn} yields: 
\begin{equation}
\label{ACTUALLL}
    \dot{\mathbf{\Phi}}_I^a
    = \mathbf{A}_{c}{\mathbf{\Phi}}_I^a+\mathbf{B}_c\hat{\mathbf{\Phi}}_{cb}^a
    \end{equation}
where ${\mathbf{\Phi}}_I^a\in \mathbb{R}^{\left(m-m_{cb}\right)\times1}$ defines actual coordination given actual boundary input $\hat{\mathbf{\Phi}}_{cb}^a\in \mathbb{R}^{m_B\times1}$ where superscript $a$ denotes actual.  Define $\mathbf{E}={\mathbf{\Phi}}_I-{\mathbf{\Phi}}_I^a$ and $\mathbf{U}=\hat{\mathbf{\Phi}}_{cb}-\hat{\mathbf{\Phi}}_{cb}^a$. By subtracting Eq. \eqref{DESIREEEE} from Eq. \eqref{ACTUALLL}, the error dynamics is obtained:
\[
\dot{\mathbf{E}}
= \mathbf{A}_{c}\mathbf{E}+\mathbf{B}_c\mathbf{U}.
\]
The paper designs macroscopic control as a linear quadratic regulator (LQR) minimizing
\[
\mathbf{J}=\int_{0}^\infty\left(\mathbf{E}\mathbf{W}_e\mathbf{E}+\mathbf{U}\mathbf{W}_u\mathbf{U}\right),
\]
where $\mathbf{W}_e\in \mathbb{R}^{\left(m-m_{cb}\right)\times \left(m-m_{cb}\right)}$ is positive semi-definite and $\mathbf{W}_u\in \mathbb{R}^{m_{cb}\times m_{cb}}$ is positive-definite. The above cost function is minimized by choosing $\mathbf{U}=-\mathbf{K}_e\mathbf{E}$, where $\mathbf{K}_E=\mathbf{W}_u^{-1}\mathbf{B}_c^T\mathbf{P}\in \mathbb{R}^{m_{cb}\times(m-m_{cb})}$ is the LQR control gain and $\mathbf{P}_e\in \mathbb{R}^{\left(m-m_{cb}\right)\times \left(m-m_{cb}\right)}$ is the solution of the algebraic Ricatti equation \cite{kalman1960contributions}:
\[
\mathbf{A}_c^T\mathbf{P}_e+\mathbf{P}_e\mathbf{A}_c-\mathbf{P}_e\mathbf{B}_c\mathbf{W}_u\mathbf{B}_c^T\mathbf{P}+\mathbf{W}_e=\mathbf{0}.
\]


\section{Microscopic Coordination}
\label{Microscopic Coordination}
First define the following position notations:
\begin{itemize}
\item $\mathbf{p}_{_{j,i}}$: Actual position of agent $j$ in cluster $i\in \mathcal{S}$
\item $\mathbf{p}_{_{d,j,i}}$: Local desired position of agent
 $j$ in cluster $i\in \mathcal{S}$
\item $\mathbf{p}_{_{RB,j,i}}$: Global desired position of agent $j$ in cluster $i\in \mathcal{S}$
\item $\mathbf{p}_{_{0,j,i}}$: Material (Relative) position of agent $j$ in cluster $i\in \mathcal{S}$
\item $\mathbf{r}_{i}$: Macroscopic desired position of cluster $i\in \mathcal{S}$
\end{itemize}

Assume $N(t)$ UAS clusters exist in $\mathcal{P}$ at time $t$. 
Existing clusters are defined by $\mathcal{S}(t)=\{1,\cdots,N(t)\}$. 
We assume that $\mathcal{V}_i=\{1,\cdots,n_i\}$ defines the UAS belonging to existing cluster $i\in \mathcal{S}$. Cluster $i$ contains either a single UAS or multiple UAS. With multiple UAS, graph $\mathcal{G}_i=\left(\mathcal{V}_i,\mathcal{E}_i\right)$ defines interagent communication in cluster $i\in \mathcal{S}$ where $\mathcal{E}_i\subset \mathcal{V}_i\times \mathcal{V}_i$ specifies edges of  of graph $\mathcal{G}_i$.

Agents (UAS) of cluster $i\in \mathcal{S}$ are considered particles of a $d_i$-D group contained by a $d_i$-D ($d_i=0,1,2,3$) simplex, called the \emph{leading simplex}. 
Define $\mathcal{V}_i=\mathcal{V}_{L,i}\bigcup \mathcal{V}_{F,i}$ where 
\begin{subequations}
\begin{equation}
i\in \mathcal{S}(t),\qquad    \mathcal{V}_{L,i}=\{1,\cdots,d_i+1\}
\end{equation}
\begin{equation}
i\in \mathcal{S}(t),\qquad     \mathcal{V}_{F,i}=\{d_i+2,\cdots,n_i\}
\end{equation}
\end{subequations}
specify identification numbers of leaders and followers, respectively. Leaders $1$ through $d_i+1$ are placed at vertices of the leading simplex. Therefore,
\begin{equation}
   \forall t,~i\in \mathcal{S},~~ \rm{rank}\left(
    \begin{bmatrix}
    \mathbf{p}_{_{2,i}}(t)-\mathbf{p}_{_{1,i}}(t)&\cdots&\mathbf{p}_{_{d_i+1,i}}(t)-\mathbf{p}_{_{1,i}}(t)
    \end{bmatrix}
    \right)
    =d_i.
\end{equation}

\textbf{Remark 3:} If $d_i=0$, cluster $i\in \mathcal{S}$ consists of a single UAS, $\mathcal{V}_i=\mathcal{V}_{L,i}$, and $\mathcal{V}_{F,i}=\emptyset$.

Assume each $d_i$-D cluster is guided by $d_i+1$ leaders placed at vertices of the leading simplex.
While leaders move independently, followers acquire desired coordination through local communication. The in-neighbor set of follower $j\in \mathcal{V}_{F,i}$ ($i\in \mathcal{S}$)
is defined by
\begin{equation}
    i\in \mathcal{S},~j\in \mathcal{V}_{F,i},\qquad \mathcal{N}_{j,i}=\left\{h\in \mathcal{V}_i\big|(h,j)\in \mathcal{E}_i\right\}.
\end{equation}
In this paper, agents are modeled as double integrators:
\begin{equation}
\label{IndDyn1}
j\in \mathcal{V}_i,\qquad
\ddot{\mathbf{p}}_{_{j,i}}=\beta_{1,i}\left(\dot{\mathbf{p}}_{_{d,j,i}}-\dot{\mathbf{p}}_{_{j,i}}\right)+\beta_{2,i}\left({\mathbf{p}}_{_{d,j,i}}-{\mathbf{p}}_{_{j,i}}\right),
\end{equation}
where $\beta_{1,i}$ and $\beta_{2,i}$ are positive control gains,
\begin{equation}
\label{IndDyn2}
j\in \mathcal{V}_i,\qquad\mathbf{p}_{_{d,j,i}}=
\begin{cases}
\rm{prescribed}&j\in \mathcal{V}_{L,i}\\
\sum_{h\in \mathcal{N}_{j,i}}w_{i,j,h}\mathbf{p}_{_{h,i}}&j\in \mathcal{V}_{F,i}\\
\end{cases}
\end{equation}
 $w_{i,j,h}>0$, and
\[
i\in \mathcal{S},~j\in \mathcal{V}_{F,i},\qquad   \sum_{h\in \mathcal{N}_j}w_{i,j,h}=1.
\]
Given communication weights, weight matrix $\mathbf{W}_i\in \mathbb{R}^{n_i\times n_i}$ is defined as follows:
\begin{equation}
\label{comweights}
    i\in \mathcal{S},\qquad w_{i,j,h}=
    \begin{cases}
    -1&j=h\wedge j\in \mathcal{V}_{i}\\
    w_{i,j,h}&h\in \mathcal{N}_{j,i}\\
    0&\rm{else}.
    \end{cases}
\end{equation}
Matrix $\mathbf{W}_i$ can be partitioned as
\begin{equation}
i\in \mathcal{S},\qquad \mathbf{W}_i=
\begin{bmatrix}
-\mathbf{I}_{d_i+1}&\mathbf{0}\\
\mathbf{\Omega}_i&\mathbf{L}_i
\end{bmatrix}
,
\end{equation}
where $\mathbf{I}_{d_i+1}\in \mathbb{R}^{\left(d_i+1\right)\times \left(d_i+1\right)}$ is the identity matrix.
\begin{theorem}\label{thm1}
If graph $\mathcal{G}_i$ defining inter-agent communication in cluster $i\in \mathcal{S}$ contains a spanning tree, and communication weights are all positive and defined as given in Eq. \eqref{comweights}, then matrices $\mathbf{L}_i\in \mathbb{R}^{(n_i-d_i-1)\times (n_i-d_i-1)}$ and $\mathbf{W}_i$ are Hurwitz.
\end{theorem}

\subsection{Microscopic Coordination and Control}
Global desired position $\mathbf{p}_{_{RB,j,i}}$ is defined by the following affine transformation:
\begin{equation}
\label{affinetransformation}
 i\in \mathcal{S},~j\in \mathcal{V}_i,\qquad \mathbf{p}_{_{RB,j,i}}=\mathbf{Q}_i(t)\mathbf{p}_{_{0,j,i}}+\mathbf{r}_i(t),
\end{equation}
where $\mathbf{Q}_i\in \mathbb{R}^{3\times 3}$ is an orthogonal rotation matrix.
Agents' material (relative) position $\mathbf{p}_{_{0,j,i}}$ ($j\in \mathcal{V}_i,~i\in \mathcal{S}$) is assumed constant.
Note that $\mathbf{p}_{_{\rm{RB},j,i}}=\mathbf{\mathbf{p}_{_{d,j,i}}}$ if agent $j\in \mathcal{V}_{L,i}$ is a leader. However, $\mathbf{p}_{_{\rm{RB},j,i}}$ differs from the local desired position $\mathbf{p}_{_{d,j,i}}$ expressed as the convex combination of in-neighbor agents per Eq. \eqref{IndDyn2} if agent $j\in \mathcal{V}_{F,i}$ is a follower.  Furthermore, macroscopic desired position $\mathbf{r}_i$ is assigned at the macroscopic level and updated by Eq. \eqref{goveqdyn}:
\begin{equation}
\label{reftraj}
    t\in [t_{0,i},t_{f,i}],~i\in \mathcal{S},\qquad \dot{\mathbf{r}}_i=K(\mathbf{r}_i,t)\bigtriangledown\Phi_a\left(\mathbf{r}_i,t\right),
\end{equation}
subject to the arrival condition $
\mathbf{r}_i(t_{0,i})=\mathbf{r}_{0,i}\in  \partial \mathcal{P}$. Note that $t_{0,i}$ is the time at which cluster $i\in \mathcal{S}$ crosses border $\partial \mathcal{P}$ to enter the planned airspace at $\mathbf{r}_{0,i}\in \partial \mathcal{P}$, and $t_{f,i}$ is the time at which cluster $i$ leaves the planned airspace at border $\mathbf{r}_{f,i}\in \partial \mathcal{P}$, e.g. $\mathbf{r}_{f,i}=\mathbf{r}_i(t_{f,i})$ ($i\in \mathcal{S}$). Given $t_{0,i}$ and $\mathbf{r}_{0,i}$ reference desired trajectory $\mathbf{r}_i(t)$ is assigned over $t\in [t_{0,i},t_{f,i}]$.

\textbf{Remark 4:} Given arrival condition $\mathbf{r}_i(t_{0,i})=\mathbf{r}_{0,i}\in  \partial \mathcal{P}$, desired reference trajectory $\mathbf{r}_i(t)$ ($t\in [t_{0,i},t_{f,i}]$) is unique.

\underline{\textit{Cluster Rotation Matrix $\mathbf{Q}_i$:}} Agents of cluster $i\in \mathcal{S}$ are treated as particles of a virtual rigid body (VRB) with a desired evolution given by rigid body motion kinematics. Defining a body frame with orthogonal unit bases $\mathbf{i}_{m,i}$, $\mathbf{j}_{m,i}$, and $\mathbf{k}_{m,i}$, $\mathbf{i}_{m,i}$, $\mathbf{j}_{m,i}$, and $\mathbf{k}_{m,i}$ are related to ground coordinate bases $\mathbf{e}_{x}$, $\mathbf{e}_{y}$, and $\mathbf{e}_{z}$ by
\[
i\in \mathcal{S},~j\in \mathcal{V}_i,\qquad
\begin{bmatrix}
\mathbf{i}_{m,i}(t)\\
\mathbf{j}_{m,i}(t)\\
\mathbf{k}_{m,i}(t)\\
\end{bmatrix}
=\mathbf{Q}_i(t)
\begin{bmatrix}
\mathbf{e}_{x}\\
\mathbf{e}_{y}\\
\mathbf{e}_{z}\\
\end{bmatrix}
\]
where
\begin{equation}
    \mathbf{Q}_{i}(t)=\begin{bmatrix}
    \cos\theta_{1,i}\cos\theta_{2,i}&\cos\theta_{1,i}\sin\theta_{2,i}&-\sin\theta_{1,i}\\
    -\sin\theta_{2,i}&\cos\theta_{2,i}&0\\
     \sin\theta_{1,i}\cos\theta_{2,i}&\sin\theta_{1,i}\sin\theta_{2,i}&\cos\theta_{1,i}\\
    \end{bmatrix}
\end{equation}
is an orthogonal matrix and $\theta_{1,i}$ and $\theta_{2,i}$ are independent rotation angles. 
This paper assumes that $\mathbf{i}_{m,i}$ is along the macroscopic desired velocity $\dot{\mathbf{r}}_i$ at any time $t$. Therefore,
\[
\dot{\mathbf{r}}_i=\|\dot{\mathbf{r}}_i\|\left(\cos\theta_{1,i}\cos\theta_{2,i}\hat{\mathbf{e}}_x+ \cos\theta_{1,i}\sin\theta_{2,i}\hat{\mathbf{e}}_y-\sin\theta_{1,i}\hat{\mathbf{e}}_z\right).
\]
Note that $\dot{\mathbf{r}}_i=\mathbf{V}(\mathbf{r}_i,t)$ has been assigned at the macroscopic level per Eq. \ref{referencedesiredvelocity}. 
Given reference velocity $\dot{\mathbf{r}}_i$, rotation angles $\theta_{1,i}$ and $\theta_{2,i}$ are obtained as follows:
\begin{subequations}
\begin{equation}
i\in \mathcal{S},\qquad     \theta_{1,i}=-\sin^{-1}\left(\frac{\dot{\mathbf{r}}_i\cdot \hat{\mathbf{e}}_z}{\|\dot{\mathbf{r}}_i\|}\right)
\end{equation}
\begin{equation}
i\in \mathcal{S},\qquad     \theta_{2,i}=\tan^{-1}\left(\frac{\dot{\mathbf{r}}_i\cdot \hat{\mathbf{e}}_y}{\dot{\mathbf{r}}_i\cdot \hat{\mathbf{e}}_x}\right).
\end{equation}
\end{subequations}

\begin{figure}[ht]
\center
\includegraphics[width=3.3 in]{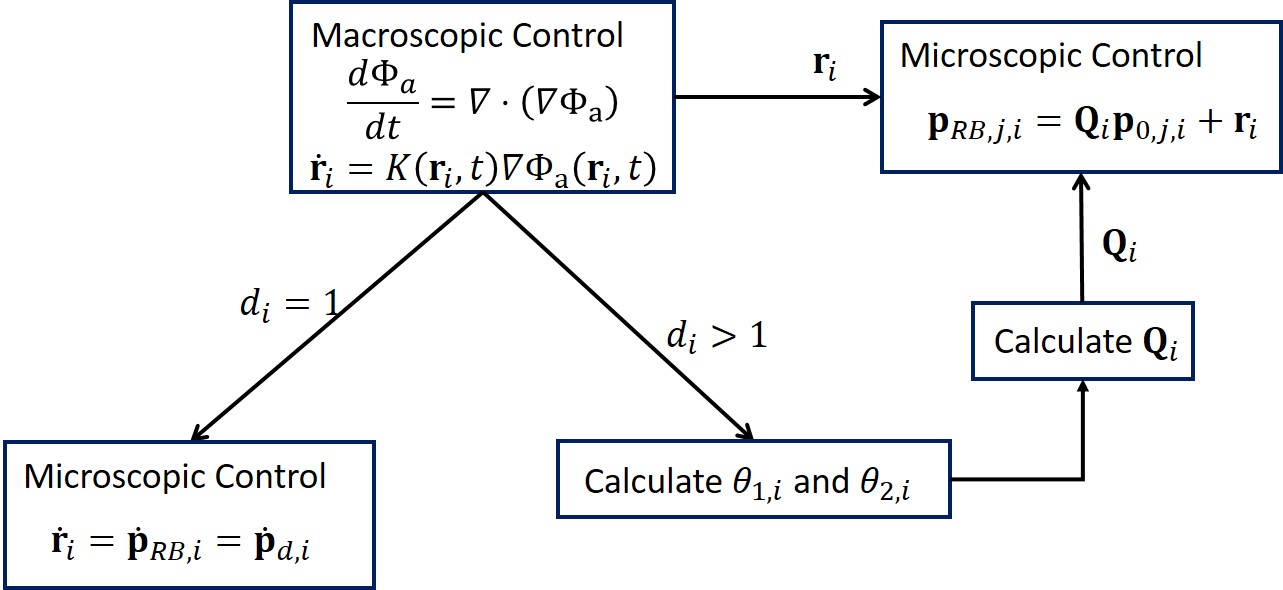}
\caption{LAUC highlighting microscopic coordination control of cluster $i\in \mathcal{S}$ .}
\label{CompleteMotionSpace}
\end{figure}

\subsection{Microscopic Collective Dynamics}

Let $\mathbf{p}_{j,i}=[p_{x,j,i}~p_{y,j,i}~p_{z,j,i}]^T\in \mathbb{R}^{3\times 1}$ denote actual position of agent $j\in \mathcal{V}_i$.
Let $\mathbf{p}_{\rm{RB},j,i}=[x_{\rm{RB},j,i}~y_{\rm{RB},j,i}~z_{\rm{RB},j,i}]^T\in \mathbb{R}^{3\times 1}$ denote global desired position of agent $j\in \mathcal{V}_i$. 
Define 

\begin{subequations}
\begin{equation}
    i\in \mathcal{S},\qquad \mathbf{P}_{\rm{RB},L,i}=
    \begin{bmatrix}
    \mathbf{p}_{_{\rm{RB},1,i}}&\cdots& \mathbf{p}_{_{\rm{RB},d_i+1,i}}
    \end{bmatrix}
    \in \mathbb{R}^{3\times (d_i+1)}
\end{equation}
\begin{equation}
    i\in \mathcal{S},\qquad \mathbf{P}_{i}=
    \begin{bmatrix}
    \mathbf{p}_{_{1,i}}&\cdots& \mathbf{p}_{_{n_i,i}}
    \end{bmatrix}
    \in \mathbb{R}^{3\times n_i}
\end{equation}
\end{subequations}
Assuming agent $j\in \mathcal{V}_i$  updates its position according to the dynamics in Eq. \eqref{IndDyn1}, collective dynamics of cluster $i\in \mathcal{S}$ is given by
\begin{equation}
i\in \mathcal{S},\qquad    \dot{\mathbf{X}}_{\rm{SYS},i}=\mathbf{A}_{\rm{SYS},i}{\mathbf{X}}_{\rm{SYS},i}+\mathbf{B}_{\rm{SYS},i}{\mathbf{U}}_{\rm{SYS},i}
\end{equation}
where 
\[
i\in \mathcal{S},\qquad
\mathbf{X}_{\rm{SYS},i}=
\begin{bmatrix}
 \rm{vec}\left({\mathbf{P}}_{i}^T\right)\\
\frac{d\rm{vec}\left({\mathbf{P}}_{i}^T\right)}{ dt}\\
\end{bmatrix}\in \mathbb{R}^{6n_i\times 1},
\]
\[
i\in \mathcal{S},~~
\mathbf{U}_{\rm{SYS},i}=\beta_{1,i} \frac{d\rm{vec}\left({\mathbf{P}}_{i}^T\right)} { dt}+\beta_{2,i} \rm{vec}\left(\dot{\mathbf{P}}_{\rm{RB},L,i}^T\right)
\in \mathbb{R}^{3(d_i+1)\times 1},
\]
\[
i\in \mathcal{S},\qquad\mathbf{A}_{\rm{SYS},i}=\mathbf{I}_3\otimes 
\begin{bmatrix}
\mathbf{0}&\mathbf{I}\\
\beta_{2,i}\mathbf{W}_i&\beta_{1,i}\mathbf{W}_i
\end{bmatrix}
\in \mathbb{R}^{6n_i\times 6n_i},
\]
\[
i\in \mathcal{S},\qquad\mathbf{B}_{\rm{SYS},i}=\mathbf{I}_3\otimes 
\begin{bmatrix}
\mathbf{0}_{n_i\times(d_i+1)}\\
\mathbf{I}_{d_i+1}\\
\mathbf{0}_{(n_i-d_i-1)\times(d_i+1)}\\
\end{bmatrix}
\in \mathbb{R}^{6n_i\times 3(d_i+1)}.
\]
Note that ``$\otimes$'' is the Kronecker product symbol, $\mathbf{0}_{n_i\times(d_i+1)}\in \mathbb{R}^{n_i\times(d_i+1)}$ and $\mathbf{0}_{(n_i-d_i-1)\times(d_i+1)}\in \mathbb{R}^{(n_i-d_i-1)\times(d_i+1)}$ are the zero-entry matrices and $\mathbf{I}_{d_i+1}\in \mathbb{R}^{(d_i+1)\times (d_i+1)}$ is the identity matrix.
Furthermore, $\rm{vec}(\cdot)$ is the matrix vectorization operator. For example,
\[
\setlength\arraycolsep{1.5pt}
\rm{vec}\left(\mathbf{P}_{i}^T\right)=
\begin{bmatrix}
p_{x,1,i}&\cdots&p_{x,n_i,i}&p_{y,1,i}&\cdots&p_{y,n_i,i}&p_{z,1,i}&\cdots&p_{z,n_i,i}
\end{bmatrix}
^T.
\]
Because $\mathbf{W}_i$ is Hurwitz per Theorem \ref{thm1}, $\mathbf{A}_{\rm{SYS},i}$ is Hurwitz as well. Therefore, the microscopic collective dynamics of cluster $i\in \mathcal{S}$ is stable.

\section{Simulation Results}
\label{Simulation Results}
A series of simulation results are presented to illustrate the proposed hierarchical coordination strategy for LAUC. First, macroscopic coordination of clustered UAS with different nominal operating speeds are shown in Section \ref{Heterogeneous Macroscopic Coordination}. Next, simulation results of  resilient macroscopic coordination are presented in Section \ref{ResilientMacroscopicCoordination1}. Section \ref{Microscopic Coordination Results} simulates microscopic coordination of an individual cluster operating with a given planned airspace region.
\begin{figure}[ht]
\center
\includegraphics[width=3 in]{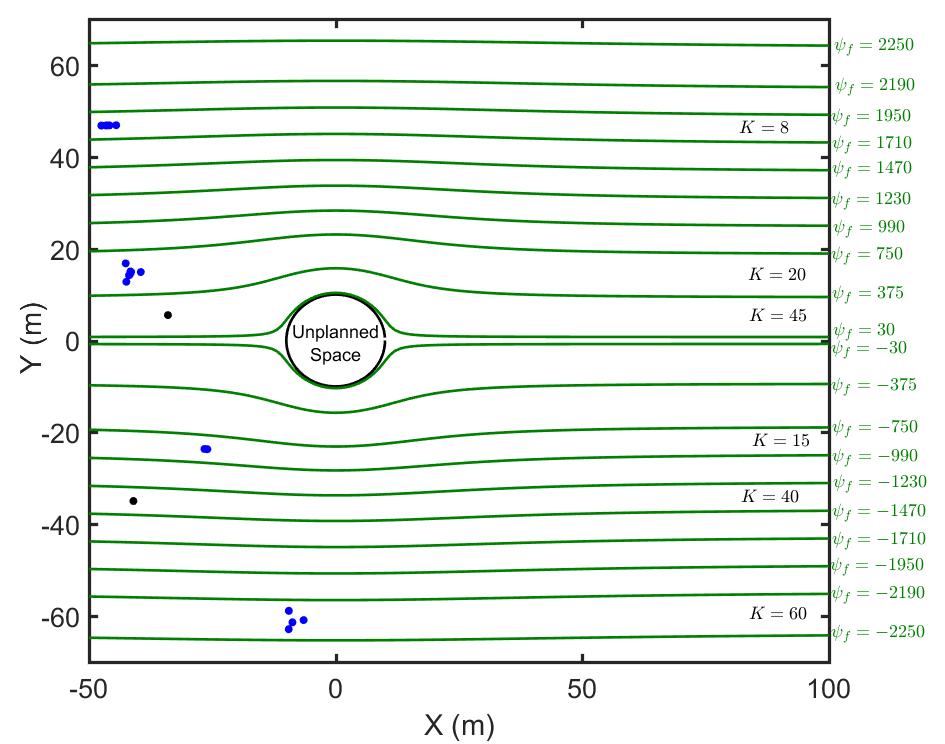}
\caption{Heterogeneous macroscopic coordination example. Level curves $\psi_f=\rm{const}$ define boundaries of the navigable channels and $\psi_f=0$ hosts a circle of radius $10 m$ excluding unplanned airspace from the planned airspace. Cost $K$ is constant along a navigable channel. }
\label{scen1heterogenmac}
\end{figure}
\subsection{Heterogeneous Macroscopic Coordination}
\label{Heterogeneous Macroscopic Coordination}
In this example, a finite airspace consisting of a single unplanned space ($n_u=1$) is considered, and the analytic approach developed in Section  \ref{Analytic Solution} is used to exclude the unplanned space. Assuming the airspace consists of a single floor ($n_s=1,~z_f=0$),  navigable channels are formed by combining a doublet flow and a uniform flow. Therefore, Eq. \eqref{PSIFFF} defines  navigable channel geometry where $\Psi_f=\rm{const}$ specifies navigable channel boundaries. In Fig. \ref{scen1heterogenmac}, naviagble channel boundaries are shown by green. Choosing $u_{\infty}=40$ and $\Delta=4000$, $\psi_f=0$ is a closed circle of radius $20m$ excluding the unplanned airspace.

\begin{figure}[ht]
\center
\includegraphics[width=2.5 in]{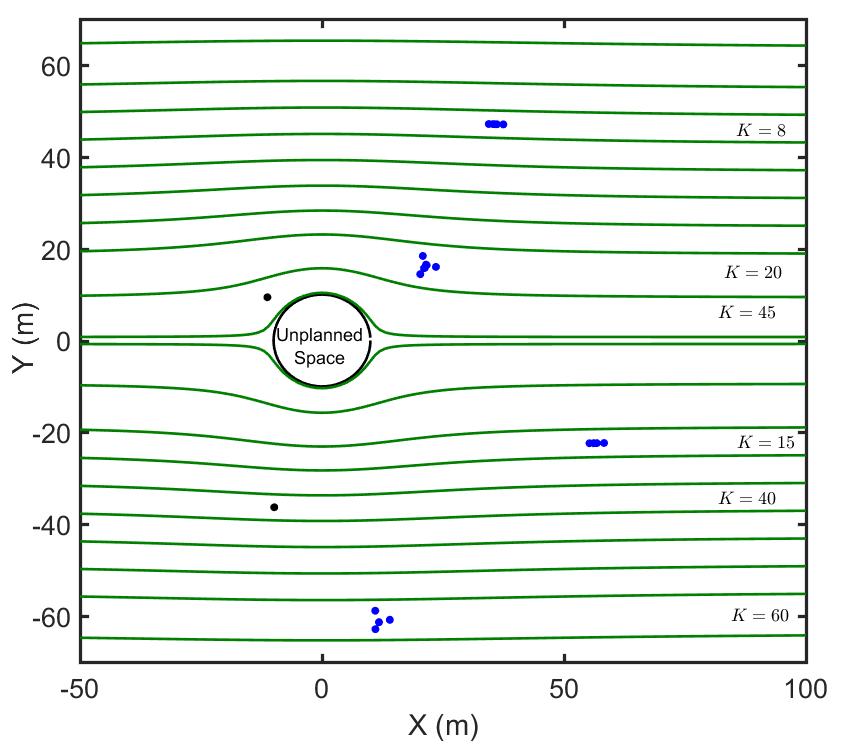}
\caption{UAS clusters moving through macroscopic channels at time $t=30s$.}
\label{scen1heterogenmac30}
\end{figure}

In this example, it is desired that nominal speed remains constant along every channel. Therefore, cost function $K=K_f$ ($f=1$) remains constant along every channel. 
To deal with UAS heterogeneity at the microscopic scale, UAS entering the airspace are clustered into $6$ groups based on their nominal speed. Assuming nominal coordination and speed are admitted by the existing UAS in the planned space, Fig. \ref {scen1heterogenmac30} shows  the six UAS clusters at time $t=30s$.

\subsection{Resilient Macroscopic Coordination}
\label{ResilientMacroscopicCoordination1}
\begin{figure}[ht]
\centering
\subfigure[$t=0s$]{\includegraphics[width=0.45\linewidth]{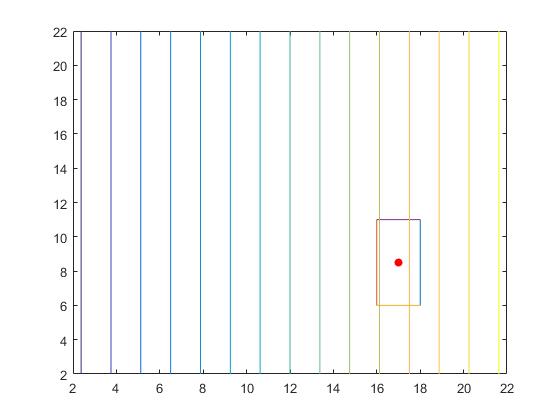}}
\subfigure[$t=10s$]{\includegraphics[width=0.45\linewidth]{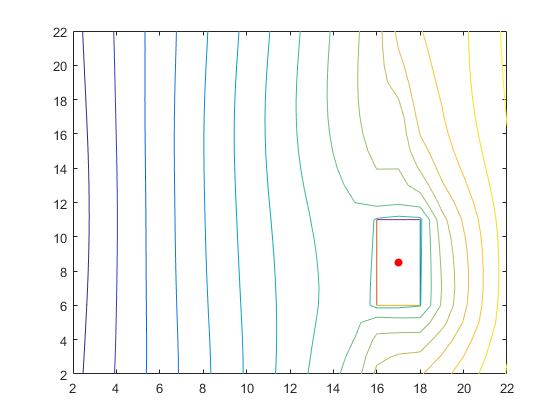}}
\subfigure[$t=20s$]{\includegraphics[width=0.45\linewidth]{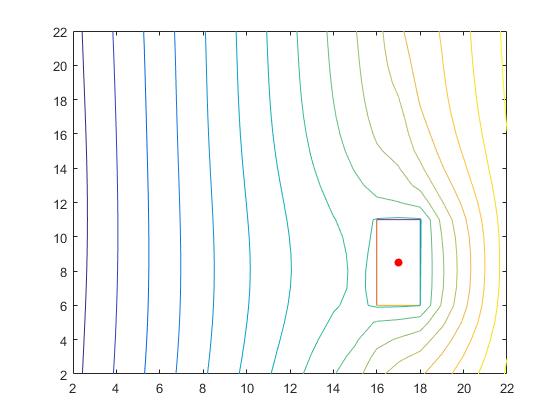}}
\subfigure[$t=50$]{\includegraphics[width=0.45\linewidth]{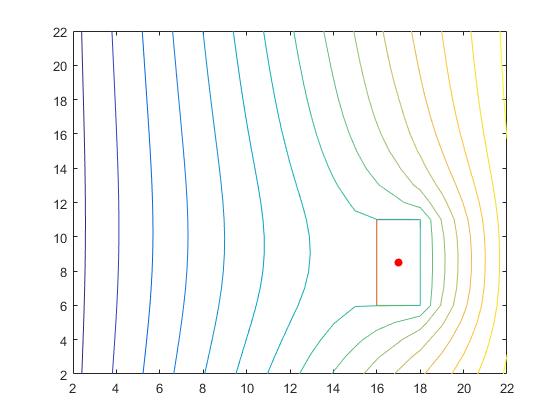}}
\caption{ Resilience of macroscopic coordination for UAS failure.  (a) A failed UAS is diagnosed at time $t=0$. The red asterisk depicts the failed UAS which must be isolated in a rectangular unplanned airspace region. (b-d) Safe navigable channels are recovered over time.
}
 \label{failure}
\end{figure}

An example of resilient macroscopic coordination  is shown in Fig. \ref{failure}. We assume that a UAS reports a failure (emergency) while moving along a safe navigable zone at $t=0$; note the failed UAS is shown by a red circle in Fig.  \ref{failure} (a)). Afterward, the failed UAS is enclosed by a rectangular unplanned airspace $\mathcal{U}_{D}$  per Fig. \ref{failure} (a). Given the geometry of the unplanned airspace, the numerical method developed in Section \ref{Nominal Velocity Assignment} is used to update the desired macroscopic coordination as the solution of Eq. \eqref{DESIREEEE}. Then, we design an LQR boundary control law to redefine the navigable channels so that the unplanned airspace is safely excluded. Hence, $\mathbf{U}=-\mathbf{K}_e\mathbf{E}$ is chosen, where $\mathbf{K}_e$ is the LQR gain obtained as the solution of the algebraic Ricatti equation \cite{bender1985linear}. Recovery of safe navigation zones is illustrated in Fig. \ref{failure} (b-d). This example illustrates how a pop-up unplanned airspace obstacle can be handled.  

\subsection{Microscopic Coordination Results}
\label{Microscopic Coordination Results}
\begin{figure}[ht]
\center
\includegraphics[width=2.2in]{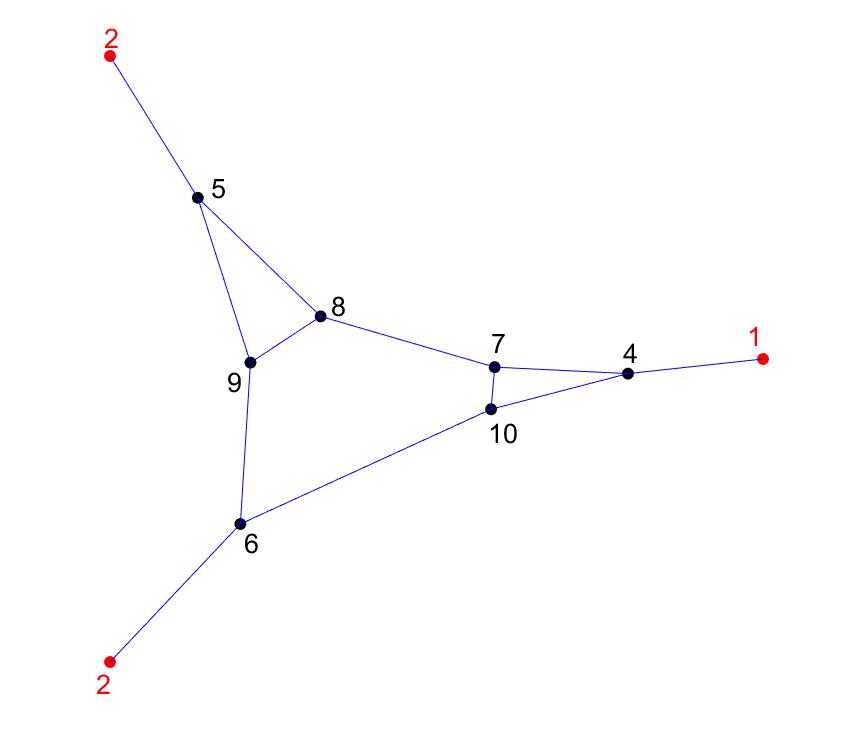}
\caption{Graph defining inter-agent communication in cluster $i$. Cluster $i$ is a $2$-D cluster ($d_i=2$); therefore, it is guided by three leaders forming a triangle. }
\label{comgraph}
\end{figure}

As a third example, consider the microscopic coordination of cluster $i$ consisting of $n_i=10$ agents. Cluster $i$  treats each UAS as a particle of a $2$-D virtual rigid body (VRB). Three leaders with indices $1$, $2$, and $3$ guide the collective motion of the cluster. The remaining agents with indices $4$ through $10$ are followers acquiring the desired coordination through local communication. The graph shown in Fig. \ref{comgraph} defines inter-agent communication in cluster $i$. Followers' communication weights are listed in Table \ref{Table2222}.
\begin{table}[t]
\caption{Communication weights $w _{i,j,h_1}$, $w _{i,j,h_2}$, and $w _{i,j,h_3}$ ($i\in \mathcal{S},~j\in \mathcal{V}_{F,i},~\mathcal{N}_{j,i}=\{h_1,h_2,h_3\}$).}
\label{Table2222}
\begin{center}
\begin{tabular}{c l l l l l l }
\hline
$j$ & $h_1$&$h_2$&$h_3$& $w _{i,j,{h_1}}$&$w_{i,j,{h_2}}$&$w _{i,j,{h_{3}}}$ \\
\hline
$4$&$1$& $7$&$10$&	$0.50$&	 $0.25$&	$0.25$\\
$5$&$2$&	$8$&	$9$&	$0.50$&	    $0.25$&	    $0.25$\\
$6$&	$3$&	$9$&	$10$&	$0.50$&	    $0.25$&	    $0.25$\\
$7$&	$4$& $8$ &$10$&	$0.40$&	    $0.30$&	    $0.40$\\
$8$&$5$& $7$ &$9$&	$0.29$&	    $0.35$&	    $0.36$\\
$9$&$5$ &$6$& $8$&	$0.31$&	    $0.4$&	    $0.29$\\
$10$&	$4$ &$6$ &$7$&	$0.45$&	    $0.25$&	    $0.30$\\
\hline
\end{tabular}
\end{center}
\end{table}
\begin{figure}[ht]
\centering
\subfigure{\includegraphics[width=0.4\linewidth]{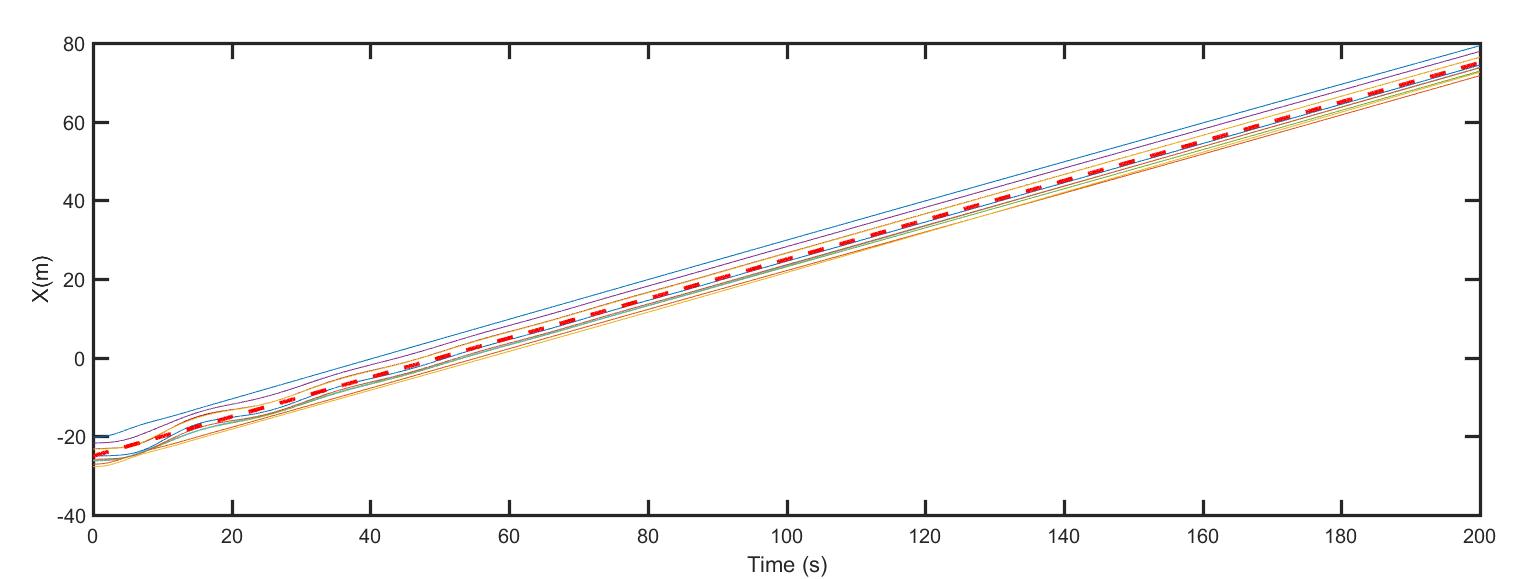}}
\subfigure{\includegraphics[width=0.4\linewidth]{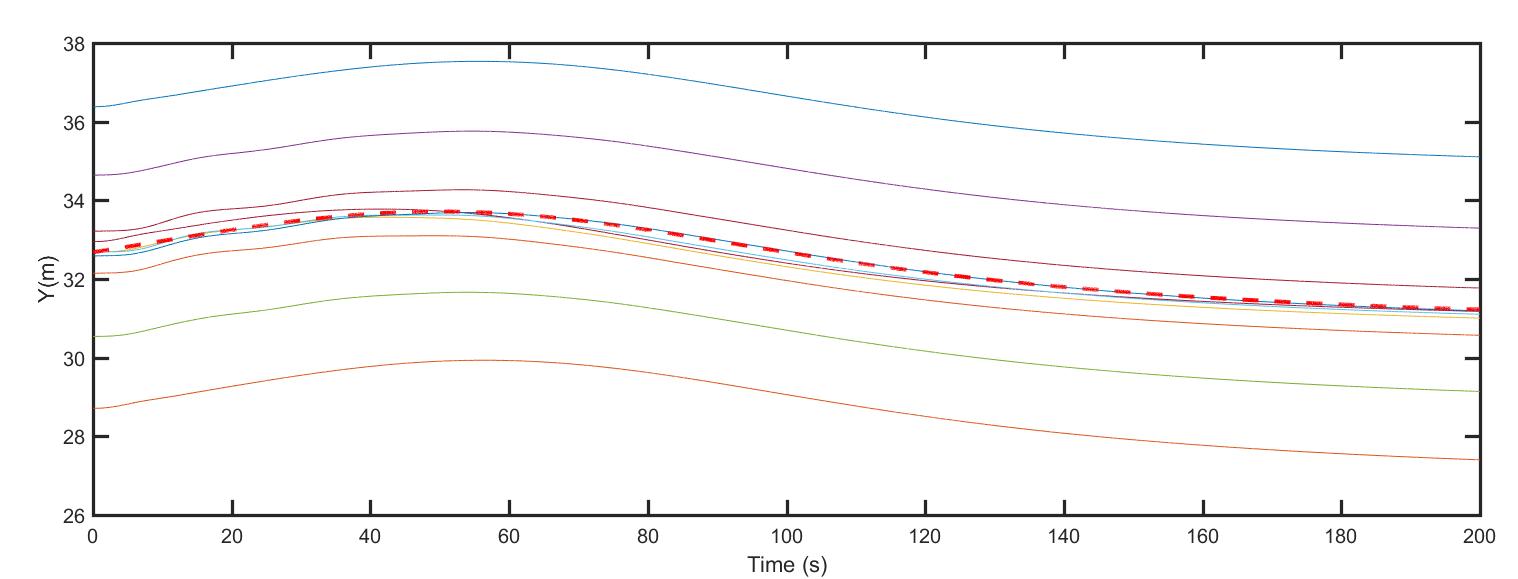}}
\subfigure{\includegraphics[width=0.4\linewidth]{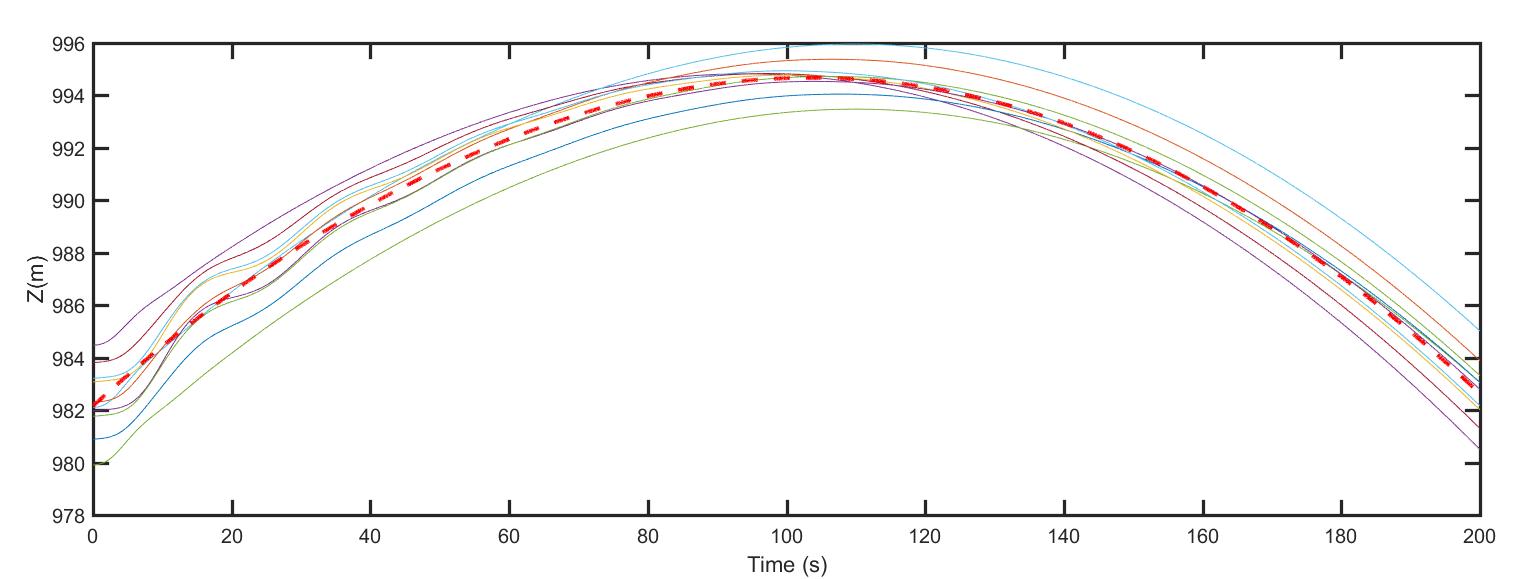}}
\subfigure{\includegraphics[width=0.38\linewidth]{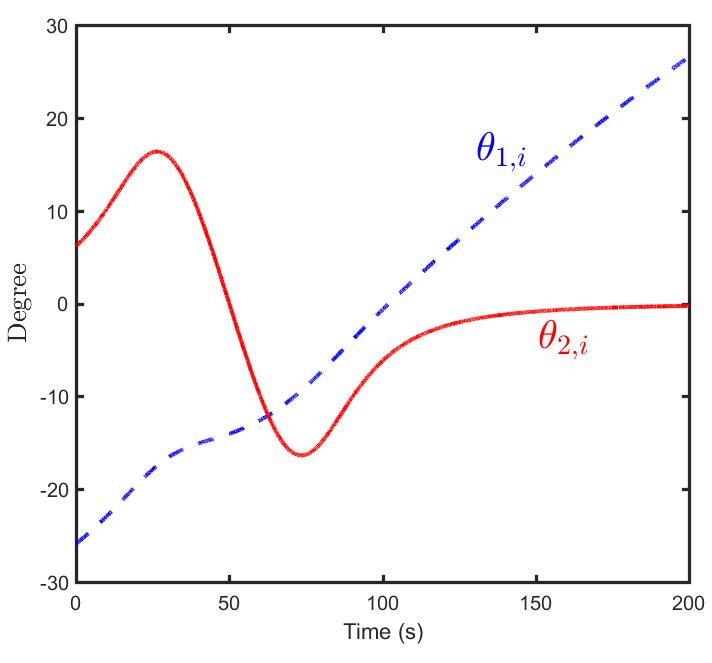}}
\subfigure{\includegraphics[width=0.8\linewidth]{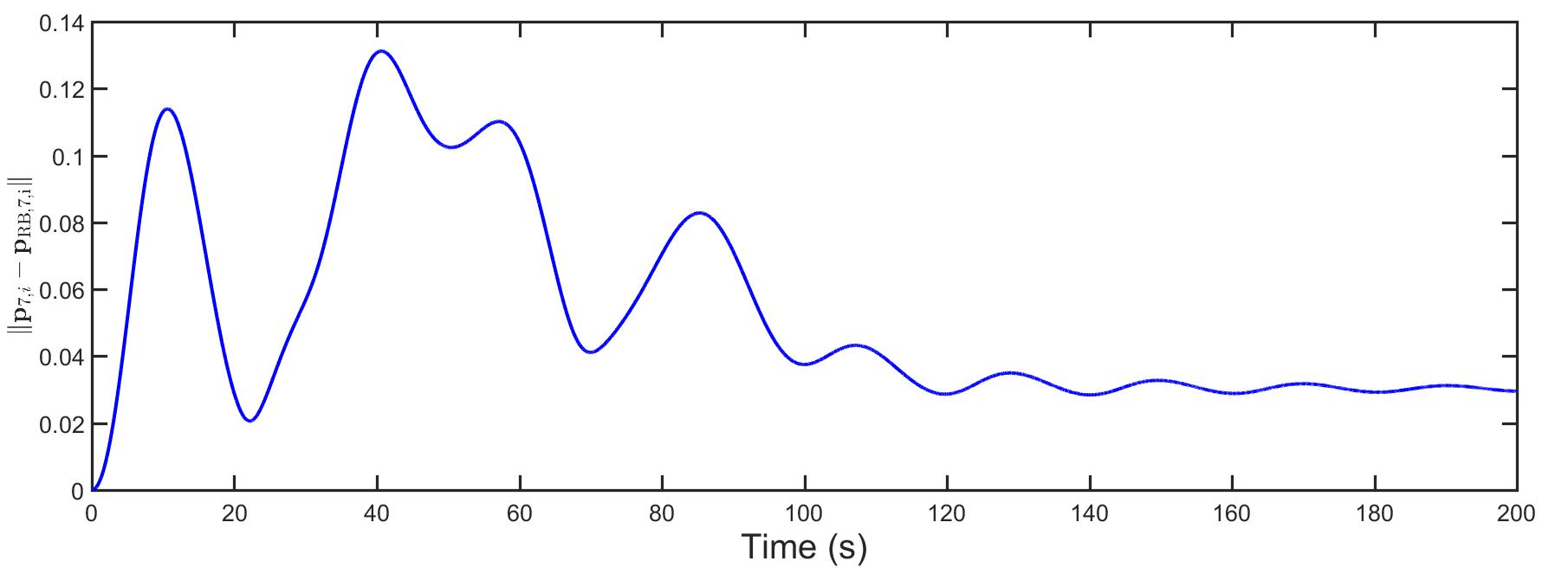}}
\caption{(a-c) $x$, $y$, and $z$ components of actual trajectory $p_{z,j,i}$ versus  $t$ for every agent $j\in\mathcal{V}_i=\{1,\cdots,10\}$;  $x_i$, $y_i$, and $z_i$ versus $t$ are shown by a dashed red trend;  (d) Macroscopic coordination rotation angles $\theta_{1,i}$ and $\theta_{2,i}$. (e) Deviation of follower $7$ from its global desired position.
}
 \label{clusteriii}
\end{figure}
\begin{figure}[ht]
\center
\includegraphics[width=3.3 in]{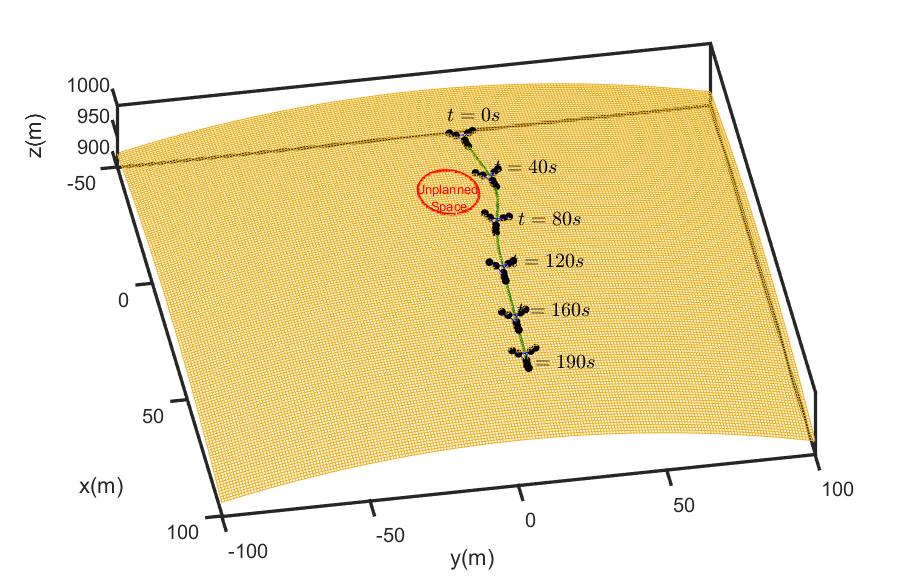}
\caption{Formation of cluster $i$ at sample times $t=0s$, $t=40s$, $t=80s$, $t=120s$, $t=160s$, and $t=190s$.}
\label{3DMASFORMATION}
\end{figure}
For this case study  cluster $i$ is instructed through macroscopic control to move on the surface 
\[
\begin{split}
f=1,\qquad    x\in &[-25,75]\\
f=1,\qquad    y\in &[-100,100]\\
f=1,\qquad z_f=&1000-5\times10^{-3}\left[\left(x-25\right)^2+y^2\right].
\end{split}
\]
Assume the single-floor airspace consists of a single unplanned space. The unplanned airspace is excluded by combining a uniform flow and a doublet flow. Therefore \[
\begin{split}
    \phi_f(x,y)=&u_{\infty,f}\left(1+\dfrac{\Delta_f}{(x-25)^2+y^2}\right)(x-25)\\
    \psi_f=&u_{\infty,f}\left(1-\dfrac{\Delta_f}{(x-25)^2+y^2}\right)y
\end{split}
\]
define the potential and stream fields over the planned airspace, respectively. It is desired that cluster $i$ moves along the path  $\mathbf{r}_i(t)=x_i(t)\hat{\mathbf{e}}_x+y_i(t)\hat{\mathbf{e}}_y+z_i(t)\hat{\mathbf{e}}_z$ where $x_i(t)$, $y_i(t)$, and $z_i(t)$ satisfy $psi_f(x_i,y_i)=375$ and $z_i=1000-5\times10^{-3}\left[\left(x_i-25\right)^2+y_i^2\right]$.
The {\color{black}macroscopic desired path $\mathbf{r}_i=x_i\hat{\mathbf{e}}_x+y_i\hat{\mathbf{e}}_y+z_i\hat{\mathbf{e}}_z$ is shown by a green curve in Fig. \ref{3DMASFORMATION}. Furthermore, components of macroscopic desired trajectory $\mathbf{r}_i$ ($x_i(t),~y_i(t),~z_i(t)$) are plotted by dashed red in Fig. \ref{clusteriii}  (a-c). Fig. \ref{clusteriii}  (a-c) also illustrates $x$, $y$, and $z$ components of the agents' actual positions versus time. Continuous curves plot actual position components $p_{x,j,i}$, $p_{y,j,i}$, and $p_{x,j,i}$ versus time for every agent $j\in \mathcal{V}_i=\{1,\cdots,10\}$.} {\color{black}Given $\mathbf{r}_{i}(t)$ macroscopic rotation angles $\theta_{1,i}(t)$ and $\theta_{2,i}(t)$ are computed at each time $t$ and plotted versus time in Fig. \ref{clusteriii}(d). Moreover, Fig. \ref{clusteriii}(e) plots deviation of follower $7$ ($\|\mathbf{p}_{7,i}-\mathbf{p}_{\rm{RB},7,i}\|$) versus time $t$.} The formation of cluster $i$ at sample times $t=0s$, $t=40s$, $t=80s$, $t=120s$, $t=160s$, and $t=190s$ are shown in Fig. \ref{3DMASFORMATION}. As illustrated, cluster $i$ successfully tracks the macroscopic desired trajectory $\mathbf{r}_i$.

\section{\hspace{0.2cm}Conclusion}
\label{Conclusion}
The proposed hierarchical LAUC strategy uses an Eulerian description of continuum mechanics to manage the space and time allocated to UAS clusters to enable heterogeneous UAS clusters to safely coordinate motion through a complex airspace sector. To this end, nominal coordination is obtained as the solution of a governing PDE with spatiotemporal parameters. Assuming airspace is finite, this paper manages the airspace capacity through controlling UAS inflow and outflow at the airspace boundaries. UAS heterogeneity is managed at both macroscopic and microscopic levels. At the macroscopic level, planned airspace can be divided into safe navigable channels and navigable channels can be allocated to different UAS classes with different nominal flight speeds. At the microscopic level, similar UAS are treated as particles of a rigid-body cluster. A leader-follower containment method is applied by each cluster to acquire the prescribed macroscopic coordination in a decentralized fashion. The paper also studies LAUC  resilience individual vehicle failure by wrapping a failed UAS with a no-fly zone defined as a "pop-up" unplanned airspace region other UAS never enter. We offer a resilient linear quadratic regulator (LQR) boundary control strategy to update the geometry of the unplanned airspace and safely plan nominal coordination in the macroscopic layer.  

This paper focuses on the efficient flow of UAS through a single LAUC airspace sector, prescribing constraints on UAS entry and exit stations based on UAS travel speed and macroscopic channel definitions.  Follow-on research is required to assure UAS will be able to safely and efficiently coordinate transitions between adjacent planned airspace sectors.

\bibliographystyle{IEEEtran}
\bibliography{reference}
 \appendices
 \section{Proofs}
 
\textbf{Proof for Theorem \ref{theorem1}:}
Let $\mathbf{A}_c$ be expressed as
\[
\mathbf{A}_c=\mathbf{G}\hat{\mathbf{A}}_c\tag{A.1}
\]
where $\mathbf{G}\in \mathbb{R}^{\left(m-m_{cb}\right)\times \left(m-m_{cb}\right)}$ is diagonal and positive definite matrix with diagonal entry
\[
\mathbf{G}_{i,i}=\sum_{j=1,~j\neq i}^{m-m_{cb}}{\mathbf{A}_c}_{_{i,j}}. \tag{A.2}
\]

The diagonal entries of matrix $\hat{\mathbf{A}}_c$ are all $-1$ while off-diagonal entries are all non-negative. Furthermore, the sums of the row entries of $\hat{\mathbf{A}}_c$ are non-negative while the sum of the entries is negative in at least one row of matrix $\hat{\mathbf{A}}_c$. Let $\hat{\mathbf{A}}_c$ be expressed as $\hat{\mathbf{A}}_c=-\mathbf{I}+\mathbf{D}$. 

The spectral radius of matrix $\mathbf{D}$, denoted $\rho(\mathbf{D})$, is less than $1$. Therefore, $-\hat{\mathbf{A}}_c$ is a non-singular M-matrix with eigenvalues located inside a disk of radius  $\rho(\mathbf{D})<1$ centered at $-1+0\mathbf{j}$. Therefore, $\hat{\mathbf{A}}_c$ and $\hat{\mathbf{A}}_c$ are both Hurwitz.

{\color{black}
 \textbf{Proof for Proposition \ref{prop1}:}
 At $\mathbf{r}\in \mathcal{U}_j$
 \[
 j=1,\cdots,n_u,\qquad q(\mathbf{r},t)=\bigtriangledown\phi_f\cdot \mathbf{n}_{u_j}=K_f\mathbf{V}_f\cdot \mathbf{n}_{u_j}=0.\tag{A.3}
 \]
 
Therefore, $\bigtriangledown \Phi$ is along the tangent unit vector $\mathbf{t}_{u_j}$ at any point $\mathbf{r}\in \mathcal{U}_j$ ($j=1,\cdots,n_u$). 
$\bigtriangledown \Psi_f(\mathbf{r},t)$ is normal to $\bigtriangledown \Phi_f(\mathbf{r},t)$ per Remark 1. 
Thus, $\bigtriangledown \Phi_f(\mathbf{r},t)$ remains $0$ along level curve
$\psi_f\left(\mathbf{r},t\right)$ which is constant at the boundary of the unplanned airspace.

\textbf{Proof for Theorem \ref{theroem2}:} Let
\[
V=\dfrac{1}{2}\int_{\mathcal{C}}\|\bigtriangledown E_f\|^2d\Omega\tag{A.4}
\]
be a Lypunpov candidate function defined over planned space $\mathcal{C}$. Then,
\[
\begin{split}
    \dot{V}=&\int_{\mathcal{C}}\dfrac{\bigtriangledown E_f}{dt}\cdot \bigtriangledown E_f dA\\
    =&-\int_{\mathcal{C}}\dfrac{d E_f}{dt} \bigtriangledown\cdot\left(\bigtriangledown E_f\right) dA+\int_{\partial \mathcal{P}}\dfrac{d E_f}{dt}  \cdot\left(\bigtriangledown E_f\cdot \mathbf{n}_p\right) ds\\+&\sum_{j=1}^{n_u}\int_{\partial \mathcal{U_j}}\dfrac{d E_f}{dt}  \cdot\left(\bigtriangledown E_f\cdot \mathbf{n}_p\right) ds.\\
\end{split}
\tag{A.5}
\]
Choosing ${U}_p$ and $U_{u_j}$ according to Eqs. \eqref{up} and \eqref{uu} and considering the error governing equation \eqref{ergoveq}, $\dot{V}$ can be converted to
\[
\label{A6}
\begin{split}
    \dot{V}=&-\int_{\mathcal{C}}\left(\dfrac{d E_f}{dt}\right)^2 dA\\
    &-\int_{\partial \mathcal{P}}\left(\bigtriangledown E_f\cdot \mathbf{n}_p\right)^2   ds-\int_{\partial \mathcal{P}}\left(\bigtriangledown E_f\cdot \mathbf{n}_p\right)^2   ds
\end{split}
.
\tag{A.6}
\]
Note that $\dot{V}$ in Eq. \eqref{A6} is negative semi-definite where $\dot{V}=0$ is the invariant set of the governing Eq. \eqref{ergoveq}. Using LaSalle’s Invariance Principle, it is concluded that $E_f(\mathbf{r},t)$ assymptotically converges to $0$ at any point $\mathbf{r}\in \mathcal{C}$ as $t\rightarrow \infty$.
}


\textbf{Proof for Theorem \ref{thm1}:}
The sum of the row entries are $0$ at row $d_i+2$ through $n_i$ of matrix $\mathbf{W}_i$ while the diagonal entries of matrix $\mathbf{L}_i$ are all $-1$. Because every leader is an in-neighbor agent to at least one follower, $d_i+1$ rows of matrix $\mathbf{\Omega}_i$ have at least one positive entry. Consequently, the sum of the row elements is negative in at least $d_i+1$ rows of matrix $\mathbf{L}_i$ while the remaining rows of matrix $\mathbf{L}_i$ are zero-sum-row. Therefore, the spectral radius of matrix $\mathbf{L}_i+\mathbf{I}_{n_i-d_i-1}$ is less that $1$ and matrix $\mathbf{L}_i$ is Hurwitz.
If $\mathbf{L}_i$ is Hurwitz, $\mathbf{W}_i$ is Hurwitz as well.

\begin{IEEEbiography}[{\includegraphics[width=1in,height=1.25in,clip,keepaspectratio]{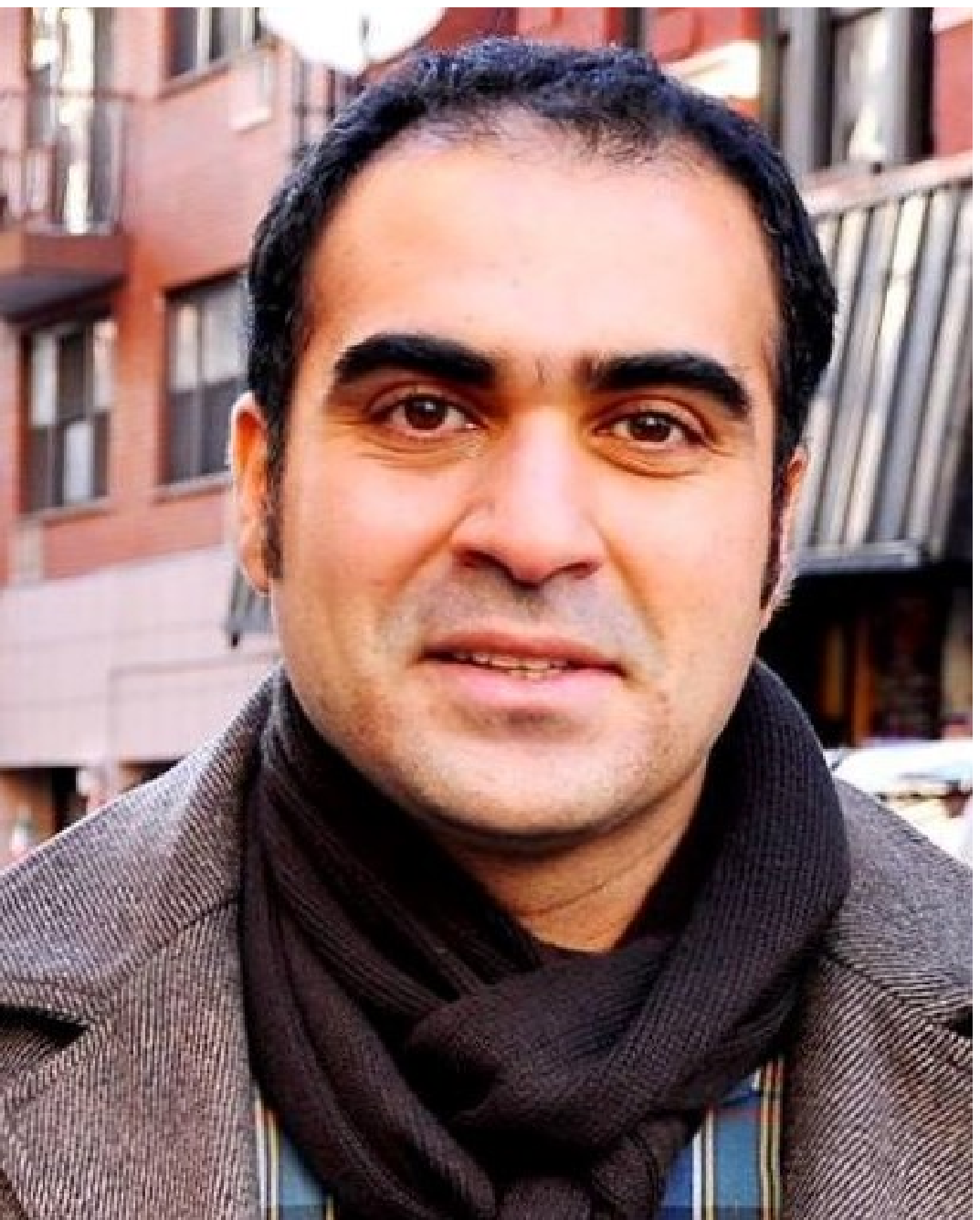}}]
{\textbf{Dr. Hossein Rastgoftar}} received the B.Sc. degree in mechanical engineering-thermo-fluids from Shiraz University, Shiraz, Iran, the M.S. degrees in mechanical systems and solid mechanics from Shiraz University and the University of Central Florida, Orlando, FL, USA, and the Ph.D. degree in mechanical engineering from Drexel University, Philadelphia, in 2015. He is currently a Postdoctoral Research Fellow at the University of Michigan, Ann Arbor, MI, USA. His current research interests include dynamics and control, multiagent systems, human-cyber physical systems, Markov decision processes, and differential game theory.
\end{IEEEbiography}
\begin{IEEEbiography}[{\includegraphics[width=1in,height=1.25in,clip,keepaspectratio]{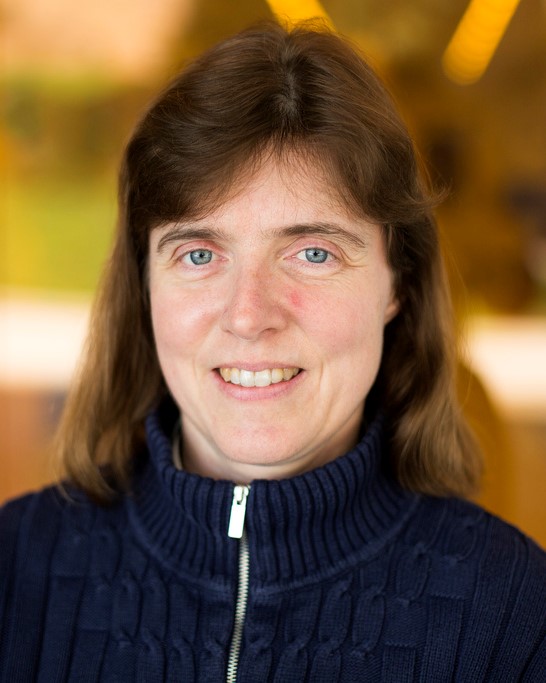}}]
{\textbf{Dr. Ella Atkins}} is a Professor of Aerospace Engineering at the University of Michigan, where she directs the Autonomous Aerospace Systems Lab and is Associate Director of Graduate Programs for the Robotics Institute.  Dr. Atkins holds B.S. and M.S. degrees in Aeronautics and Astronautics from MIT and M.S. and Ph.D. degrees in Computer Science and Engineering from the University of Michigan.  She is past-chair of the AIAA Intelligent Systems Technical Committee and has served on the National Academy's Aeronautics and Space Engineering Board, the Institute for Defense Analysis Defense Science Studies Group, and an NRC committee to develop an autonomy research agenda for civil aviation. She pursues research in Aerospace system autonomy and safety.
\end{IEEEbiography}

\end{document}